\documentclass[a4paper,12pt]{article}
\usepackage[normalem]{ulem}
\usepackage[left=2cm,right=2cm,top=2cm,bottom=1cm]{geometry}
\usepackage{epsfig,amsfonts,amssymb}
\usepackage{jheppub}
\usepackage{esvect}
\usepackage{comment}
\usepackage{amsmath, amssymb, slashed, epsf, color, graphicx, latexsym, bm}
\usepackage{mathrsfs}
\usepackage{tensor}
\usepackage{physics}
\usepackage{epsfig}
\usepackage{graphics}
\usepackage{mathtools}
\usepackage{caption}
\usepackage{float}
\usepackage{bbm}
\usepackage{subcaption}
\usepackage{enumitem}
\usepackage{verbatim}
\usepackage{blindtext}
\usepackage[mathlines]{lineno}
\usepackage{xcolor}
\usepackage{amsmath}
\usepackage{bm}
\usepackage{framed}
\usepackage{varioref}
\colorlet{shadecolor}{lightgray}
\usepackage{hyperref}
\usepackage{parskip}
\usepackage{soul}

\usepackage[utf8]{inputenc}
\usepackage{soul}

\definecolor{blue}{rgb}{0.0, 0.0, 1.0}

\usepackage{tikz}
\usetikzlibrary{arrows,snakes,backgrounds}
\usetikzlibrary{decorations.markings}
\usepackage{setspace}

\newcommand{\cO}{\mathcal{O}}
\newcommand{\rn}{r_{\mathrm{n}}}
\newcommand{\ruc}{r_{\mathrm{uc}}}
\newcommand\augniva[1]{\textcolor{blue}{(AR: #1)}}

\tikzset{middlearrow/.style={
		decoration={markings,
			mark= at position 0.5 with {\arrow[scale=2]{#1}} ,
		},
		postaction={decorate}
	}
}

\tikzset{2multiarrow/.style={
		decoration={markings,
			mark= at position 0.35 with {\arrow[scale=2]{#1}} ,
			mark= at position 0.75 with {\arrow[scale=2]{#1}} ,
		},
		postaction={decorate}
	}
}

\tikzset{3multiarrow/.style={
		decoration={markings,
			mark= at position 0.25 with {\arrow[scale=2]{#1}} ,
			mark= at position 0.5 with {\arrow[scale=2]{#1}} ,
			mark= at position 0.75 with {\arrow[scale=2]{#1}} ,
		},
		postaction={decorate}
	}
}

\tikzset{4multiarrow/.style={
		decoration={markings,
			mark= at position 0.2 with {\arrow[scale=2]{#1}} ,
			mark= at position 0.4 with {\arrow[scale=2]{#1}} ,
			mark= at position 0.6 with {\arrow[scale=2]{#1}} ,
			mark= at position 0.8 with {\arrow[scale=2]{#1}} ,
		},
		postaction={decorate}
	}
}

\begin{document}
\begin{center}
{\large \bf Logarithmic corrections to near-extremal entropy of charged de Sitter black holes}
\end{center}
\vspace*{1cm}
\centerline{\rm Sabyasachi Maulik$^{a}$, Arpita Mitra$^{b}$, Debangshu Mukherjee$^{c}$, Augniva Ray$^{d}$}
\vspace*{4.0ex}
{\small{
			\centerline{\it ~$^a$Department of Physics, Indian Institute of Technology Kanpur,}
            \centerline{\it Kalyanpur, Kanpur, Uttar Pradesh 208 016, India.}
			\centerline{\it ~$^b$Department  of  Physics,  Pohang  University  of  Science  and  Technology (POSTECH),}
            \centerline{\it Pohang  37673,  Korea.}
			\centerline{ \it ~$^c$Asia Pacific Center for Theoretical Physics, POSTECH, Pohang 37673, Korea.}
			\centerline{ \it ~$^d$Department of Physics and Astronomy \& Center for Theoretical Physics, Seoul National University,}
            \centerline{ \it Seoul 08826, Korea.}
\vspace*{0.3cm}
	\centerline{\footnotesize Email: mauliks@iitk.ac.in, arpitamitra89@gmail.com, debangshu.mukherjee@apctp.org, augniva15@snu.ac.kr}
	
	\vspace*{5.0ex}
	
\centerline{\bf Abstract} \bigskip

We calculate the logarithmic temperature corrections to the thermodynamic entropy of four-dimensional near-extremal Reissner-Nordstr\"{o}m de Sitter (dS) black hole by computing a one-loop contribution within the path integral framework in the near-horizon limit. Due to the presence of three horizons, the extremal limit of a charged dS black hole is fundamentally different from its flat and AdS counterparts. In the near-horizon limit, there are three distinct extremal limits known as cold, Nariai, and ultracold configurations. We compute the tensor zero modes of the Lichnerowicz operator acting on linearized metric perturbations for the cold and Nariai extremal limits which are associated with near-horizon AdS$_2$ and dS$_2$ asymptotic symmetries. {{In particular in the near-Nariai limit we compute the quantum corrections to the Hartle-Hawking wavefunction at late times. Our computation establishes the result that at leading order, the small temperature corrections to the extremal entropy is universal in the cold and Nariai limit, paving the way for similar such computations and tests in higher dimensional dS black hole spacetimes, including rotating dS black holes.}}

\newpage
\tableofcontents
\section{Introduction}\label{sec:intro}
Black holes possess a non-zero thermodynamic entropy which, in the semi-classical regime, is proportional to one-quarter of the area of their event horizon, measured in Planck units \cite{PhysRevD.7.2333, Hawking:1975vcx}. However, this formula receives quantum corrections, which, in the leading order, are proportional to the logarithm of the area of the black hole horizon \cite{Solodukhin:1994yz, Fursaev:1994te, Mann:1997hm, Kaul:2000kf, Carlip:2000nv, Solodukhin:2010pk, Banerjee:2010qc, Banerjee:2011jp, Sen:2012kpz, Sen:2012cj}. 
There are two distinct approaches to estimate these corrections. The first relies on a microscopic construction of the black hole, where one can estimate the entropy of the black hole \textit{exactly} from a counting of microstates, a direction first pursued successfully in the seminal work \cite{Strominger:1996sh}. Another route uses effective theories. One expects that the low-energy effective action of the UV complete quantum gravity theory is a theory of supergravity. Then, one could calculate the one-loop corrections to the classical entropy from the massless sector of supergravity theory. An agreement between the results obtained via the two methods places non-trivial tests on the low energy gravitational effective actions of UV complete quantum gravity theories \cite{Sen:2012cj, Sen:2012kpz}. Indeed, using the low energy effective supergravity actions, Sen and collaborators established \cite{Banerjee:2010qc,Banerjee:2011jp,Sen:2012kpz,Sen:2012cj,2013JHEP...04..156S} that the general form of \textit{quantum entropy} (i.e., including corrections to the semi-classical value of the entropy), is obtained as
\begin{eqnarray} \label{eq:general_expression_for_black_hole_entropy}
    S_{\textrm{quant}} =S_0 + c_{\textrm{log}} \log S_0 + \textrm{sub-leading~terms}\,, \quad S_0 = \frac{A}{4G_N}\,, 
\end{eqnarray} where $A$ and $G_N$ are respectively the area of the extremal horizon, and the Newton's constant. It was shown in \cite{Banerjee:2010qc, Banerjee:2011jp, Sen:2012kpz} that in $\mathcal{N=}2, 4, 8$ ungauged supergravities, the term $c_{\textrm{log}}$ is actually universal, i.e., it is independent of any charge of the black hole and only depends on the massless fields of the theory in consideration. Moreover, they reproduced the corresponding corrections to entropy from the statistical counting of black hole microstates. See \cite{2013JHEP...04..156S, Bhattacharyya:2012bps, Banerjee:2020bkp, Karan:2020kpx, Karan:2021kpx, Banerjee:2021bpx} for an incomplete list of subsequent works on log correction to black hole entropy. Logarithmic contribution to entropy in anti-de Sitter spacetime have been a subject of extensive study for over a decade, both in their own right and in the context of gauge/gravity duality \cite{Bhattacharyya:2012bgm, Liu:2017prl, Liu:2017lpr, Gang:2019gnp, Pando:2019tti, Benini:2020bgp, Pando:2020pzx, Hristov:2021hrx, David:2021dgn, David:2021dgl, Karan:2022ksx, Bobev:2023bdh, Karan:2024ksb}.

Although this agreement is impressive, it raises a critical question that must be addressed. The key issue is as follows: the above estimate for the black hole entropy exhibits a huge degeneracy of the black hole ground state. In the presence of supersymmetry, the degeneracy is understood because supersymmetry protects the ground state degeneracy. However, one could wonder whether such an expression for the entropy could be logically argued if supersymmetry is broken, for example in the presence of a small but non-zero temperature $T$. Stated in another way, what protects this huge ground state degeneracy if the black hole is non-extremal? The logical conclusion in such a scenario would be that there might, in fact, be large ($T$ dependent) corrections to equation \eqref{eq:general_expression_for_black_hole_entropy}. Furthermore, \cite{Preskill:1991tb} predicts the breakdown of semiclassical  approximation for a near extremal black hole below the temperature at which energy associated with the black hole and a Hawking quanta becomes equal. 

In recent years, temperature dependent quantum corrections to the right-hand side of equation \eqref{eq:general_expression_for_black_hole_entropy} were estimated \cite{Iliesiu:2020qvm, Iliesiu:2022onk, Banerjee:2023quv, Kapec:2023ruw, Rakic:2023vhv, Banerjee:2023gll, Maulik:2024dwq, Kolanowski:2024zrq, Kapec:2024zdj, Modak:2025gvp} by utilizing the near-horizon geometry of near-extremal black holes, and tools from first-order perturbation theory. These $\log T$ corrections to the black hole entropy can be obtained from the zero modes of the kinetic operator acting on field fluctuations. Moreover, it was argued that, indeed, such corrections coming from the graviton zero mode sector are \textit{universal}, independent of the parent black hole geometry. In fact, at least in the context of Kerr black holes this claim has been proven in \cite{Arnaudo:2024bbd} using conformal block techniques. 


So far, the focus of such computations has been restricted to asymptotically flat and asymptotically anti-de Sitter (AdS) black holes. Whereas the reason for such a focus could have been that the authors had string theory compactifications and supergravity solutions in mind\footnote{See \cite{Witten:2001kn, Strominger:2001pn, Goheer:2002vf,Kachru:2003aw} for discussions on the role of dS and the formulation of a `dS/CFT' correspondence, as well as \cite{Castro:2011xb, Basu:2019mjo} for directions in evaluating quantum gravity partition functions on dS. }, de Sitter (dS) black holes are equally amenable to the tools of perturbation theory and it is on them that this manuscript focuses on. De Sitter  spacetime, which closely resembles the state of our universe in about $10^{11}$ years due to accelerated expansion, is the maximally symmetric solution to the Einstein equations with a positive cosmological constant $\Lambda$. In particular, we consider an electrically charged Reissner-Nordstr\"om black hole in this background to study the universal $\log{T}$ correction to the entropy near the extremal limit. The blackening factor involved in the metric solution can have up to three real roots corresponding to: the inner (Cauchy) horizon, the event (black hole) horizon, and the cosmological horizon. The exact positions of these horizons depend on the values of mass, charge and $\Lambda$. A significant distinction between a black hole horizon and a cosmological horizon lies in the observer-dependent nature of the cosmological horizon, which defines the boundary of the causally accessible region for an observer at any time or spatial location. This horizon is a consequence of the expanding universe and different observers thus can see and influence different parts of the spacetime and experience different horizons and areas of all these horizons can themselves be interpreted as a measure of the respective entropies. 

The concept of entropy for a cosmological horizon was first proposed in \cite{Gibbons:1977mu}. For an observer following a timelike geodesic in the static patch (causally accessible region) bounded by past and future cosmological horizons-- the entropy is proportional to the area of the cosmological horizon $A_{\rm CH}$. In contrast to flat or AdS cases, an empty dS space maximizes the generalized entropy \cite{Bekenstein:1972tm} of any state within the static patch \cite{Bousso:2000nf, Bousso:2000md}. 
Therefore, dS space has a finite-dimensional Hilbert space, unlike both the flat and AdS cases. A Hilbert space with a dimension given by $e^{A_{\rm CH}/4G_N}$ up to a normalization constant is adequate to describe all possible states within the static patch including all matter, black holes and cosmological horizon since the
maximum possible entropy of a state of the static patch is $A_{\rm CH}/4G_N$. Due to the presence of these three horizons, the extremal limit for charged dS black hole has a fundamentally distinct flavour compared to its flat and AdS counterparts. There exist three extremal limits where any two among the three horizons, or all of them coalesce \cite{Romans:1991nq, Mann:1995vb, Booth:1998gf}. One of them is the cold configuration where the inner and outer horizons coincide, resulting in a near-horizon AdS$_2$ × S$^2$ geometry. Another limit arises when the outer and cosmological horizon coalesce, leading to a near-horizon geometry of dS$_2$ × S$^2$; this is usually known as the Nariai limit \cite{Nariai:1950srt, Nariai:1999iok}. When all three horizons coincide, it results in an ultracold extremal black hole with the near-horizon geometry of Mink$_2$ × S$^2$. 

In \cite{Castro:2022cuo}, 
consistent two-dimensional effective theories away from extremality are constructed from dimensional reduction on S$^2$ \cite{Nayak:2018qej} by leveraging spherical symmetry, that captures the dynamics and deformations of AdS$_2$, dS$_2$ and Mink$_2$. These two-dimensional theories lead to Jackiw-Teitelboim (JT) gravity \cite{Almheiri:2014cka, Maldacena:2019cbz, Moitra:2022glw} in AdS$_2$ and dS$_2$ cases respectively, as well as the Callan-Giddings-Harvey-Strominger (CGHS) model for Mink$_2$ case \cite{Afshar:2019axx}. The effective Schwarzian description of boundary dynamics for AdS$_2$ JT gravity gives rise to the $\log{T}$ corrections to the logarithm of the partition function \cite{Iliesiu:2020qvm}. This motivates us to study $\log{T}$ correction  for the dS black hole near extremality. While in AdS spacetime, the path integral for a single boundary can be interpreted as the computation of the boundary theory's partition function at a finite temperature, for dS case, the path integral corresponds to evaluating the wavefunction of the universe, based on the no-boundary proposal \cite{Hartle:1983ai} and involves no boundary reparametrization symmetry. To study the large gauge modes at dS$_2$ boundary one has to consider nontrivial contour in the path integral. The contour in this scenario traverses spacetime regions with different signatures \cite{Maldacena:2019cbz, Moitra:2021uiv} and one can consider two distinct contours. The first, the Hartle-Hawking contour \cite{Hartle:1983ai}, begins in a Euclidean region with signature 
$(2,0)$ corresponding to two spacelike directions, and smoothly transitions into dS space with signature 
$(1,1)$ following an analytic continuation in the radial coordinate. The second one is the Maldacena contour \cite{Maldacena:2019cbz}, which starts with a spacetime region of signature $(0,2)$, representing two timelike directions, before evolving into a dS$_2$ region. In particular we consider the Hartle-Hawking contour to evaluate the finite temperature corrections to the path integral. 
Both contours give the same universal 1-loop contribution to the path integral arising from graviton zero modes of the kinetic operator\footnote{Note that we only focus on the logarithmic corrections arising from graviton \emph{tensor} zero modes which have support only on the $\left(\tau, y\right)$ plane. Additional corrections may arise from graviton \emph{vector} zero modes supported on the $S^2$ part of the near-horizon geometry, and other fields such as a $U(1)$ gauge field in the present case. Our computation does not involve them. These issues have been addressed in recent works \cite{Blacker:2025zca,Shi:2025amq}.}

Before we end the introduction of an article concerning itself with the calculation of `corrections' to dS entropy, let us, in passing, acknowledge the elephant in the room. The statistical interpretation of this entropy, let aside corrections thereof, still remains open to considerations. In other words, it is far less clear what this entropy is `counting', if anything at all. As the horizon itself is observer dependant, it is unclear where the microstates that give rise to this entropy lie. Furthermore, since the Hamiltonian vanishes on dS, it is unclear why the partition function, now a trivial sum over identity states, should carry any non-trivial information at all. These observations make entropy in dS spacetimes fundamentally different from the corresponding notion in AdS spacetimes. In fact, these considerations led the authors of \cite{Bobev:2022lcc} to put forward the idea that on dS, the exponential of the entropy is not an integer, and therefore, quite surprisingly, entropy at least in the context of dS spacetimes does \textit{not} arise from counting microscopic degrees of freedom. However, they have shown that at least $4d$ dS spacetime, as solutions of four-derivative  purely gravitational theory, can be embedded in EAdS $\times$ S$^7/\mathbb{Z}_k$ solution of M theory, which then allows the authors to \textit{associate} the gravitational entropy with the partition function of the dual SCFT on S$^3$. In the absence of a general statistical interpretation, such directions might pave the way for a rigorous interpretation of the dS entropy. However, in this article, we will have little occasion to pursue such directions, and will be willfully agnostic to the statistical origin of either the entropy itself or its corrections.

The rest of the article is organized as follows: we briefly review electrically charged RN-dS$_4$ black holes in section \ref{sec:sec2}, highlighting their three extremal limits. In Section \ref{sec:extremalsec}, we discuss in detail the near-horizon geometry of the black hole in both extremal and near-extremal limits. Section \ref{sec:sec3} is dedicated to the computation of graviton zero modes for the three near-extremal, near-horizon backgrounds. For cold and Nariai limits, away from extremality these zero modes ultimately produce the coveted $\log T$ corrections, which are discussed in Section \ref{sec:sec4}. We end this article in Section \ref{sec:conclude} with a summary of the results and a brief mention of some potential directions for future research. In the appendix \ref{appendix:A}, we point out the power series expansion in temperature for the ultracold case in contrast to the other two extremal limits. We have computed the on-shell action in the Nariai limit in \ref{appendix:B}.

\section{Reissner-Nordstr\"{o}m \MakeLowercase{d}S$_4$ black holes}\label{sec:sec2}

In this section, we review a few vital properties of Reissner-Nordstr\"{o}m black holes in $\left(3+1\right)$ dimensional de Sitter spacetime (henceforth referred to as RN-dS$_4$ black holes), focussing on their near-extremal, near-horizon geometries. These properties are essential for the subsequent analysis of quantum corrections to black hole entropy. 
For detailed discussions on aspects of RN-dS$_4$ black holes the reader may refer to \cite{Romans:1991nq, Mann:1995vb, Booth:1998gf, Astefanesei:2003gw} or more recent commentaries in \cite{Montero:2019ekk, Castro:2022cuo, Aalsma:2023mkz}. We primarily follow the conventions set in \cite{Castro:2022cuo}.

Our starting point is the Einstein-Hilbert action in $\left(3+1 \right)$ spacetime dimensions with a positive cosmological constant, and a minimally coupled $U(1)$ gauge field. The action is given by\footnote{For the sake of completeness, we have an explicit factor of the gravitational coupling $G_N$ here but subsequently, we will take $G_N=1$.}
\begin{align}\label{eq:EHM_action}
    I^{(4D)} = \frac{1}{16\pi G_N}\int d^4x \sqrt{-g}\, \left(R - 2\Lambda_4 - \tensor{F}{_{\mu\nu}}\tensor{F}{^{\mu\nu}}\right)\,,
\end{align}
where the positive cosmological constant $\Lambda_4$ is related to the radius of dS$_4$ as usual by: $\Lambda_4= \frac{3}{\ell_4^2} > 0$. 
The variation of the action with respect to metric and gauge field gives us the following equations of motion
\begin{equation}\label{eq:EOM}
    R_{\mu\nu}-\frac{1}{2}g_{\mu\nu}R+\Lambda_4 g_{\mu\nu}=2F^{\rho}_{\ \ \mu}F_{\rho\nu}-\frac{1}{2}g_{\mu\nu}F^2\,,\quad  \nabla_{\mu}F^{\mu\nu}=0\,.
\end{equation}
An \emph{electrically charged} RN-dS$_4$ black hole is a spherically symmetric solution of 
\eqref{eq:EOM}, characterized by three independent constants $M$, $Q$ and $\ell_4$ and is given by
\begin{equation}\label{eq:RNdS4_solution}
    \begin{split}
    ds^2 &= -f(r)dt^2+{1\over f(r)}dr^2+r^2 \left(d\theta^2 + \sin^2\theta d\phi^2\right)\,,\\
    A &= -\frac{Q}{r_{+}} \left(1-\frac{r_{+}}{r}\right)\,dt,
    \end{split}
\end{equation}
where the blackening factor $f(r)$ is
\begin{equation}\label{eq:blackening_func}
    f(r)=1-{2M\over r}+{Q^2\over r^2}-{r^2\over \ell_4^2}\,.
\end{equation}
The constants $M$ and $Q$ are the \emph{mass} and \emph{charge} of the black hole, respectively. The constant $r_{+}$ (which we define shortly) denotes the radius of the outer event horizon of the black hole. The blackening factor \eqref{eq:blackening_func} determines the horizons of the black hole. Being a quartic polynomial in $r$, it has four roots. But even when all the roots are real, only three of them are positive. They correspond to three \emph{physical} horizons traditionally known as the inner horizon ($r_{-}$), the outer horizon ($r_{+}$), and the cosmological horizon ($r_c$) with the hierarchy $r_{-} < r_{+} < r_{c}$. The presence of the cosmological horizon is a distinctive feature of dS 
black holes, otherwise absent in black holes in flat or AdS spacetimes. 

Given the intricate structure of the blackening factor, our goal in this section is to review the various near-extremal, near-horizon geometries emerging out of RN-dS$_4$ black holes. To consider the extremal limit, it is useful to express the blackening factor in terms of the locations of the three horizons mentioned above as 
\begin{align}
     f(r) = -\frac{1}{r^2\ell_4^2}\left(r-r_{-}\right)\left(r-r_{+}\right)\left(r-r_{c}\right)\left(r+r_{-}+r_{+}+r_{c}\right)\,.
\end{align}
By comparing \eqref{eq:blackening_func} with the factored form of $f(r)$, we derive expressions for the two conserved charges $M$ and $Q$ 
\begin{subequations}
\begin{align}
   M &= \frac{\left(r_{-}+r_{+}\right) \left(r_{+}+r_{c}\right) \left(r_{-}+r_{c}\right)}{2\ell_4^2}\,,\\
    Q &= \frac{\sqrt{r_{-}r_{+}r_{c}\left(r_{-}+r_{+}+r_{c}\right)}}{\ell_4}\,.
\end{align}
\end{subequations}
The presence of three different length scales in the theory leads to an additional relation among the three roots and the curvature radius $\ell_4$, 
\begin{equation}\label{eq:curvature_and_rc}
     \ell_4^2 = r_-^2 + r_+^2 + r_c^2 + r_-r_+ + r_-r_c + r_+r_c\,.
\end{equation}
Demanding the absence of a naked singularity puts a constraint on the $(M,Q)$ parameter space, and the constraint is determined by the positivity of the  discriminant of $f(r)$:
\begin{equation}
    \mathrm{Discr}_{4} = \frac{16}{\ell_4^6} \left(\ell_4^4 \left(M^2-Q^2\right) + \ell_4^2 \left( 36 M^2 Q^2 -27 M^4 - 8 Q^4\right) - 16 Q^6\right)\,.
\end{equation}
Requiring positivity of the discriminant and the mass i.e. demanding Discr$_{4} > 0$ and $M>0$  is sufficient for the reality and positivity of the roots of $f(r)$. The valid parameter regime for RN-dS$_4$ black hole in the $(M,Q)$ space is illustrated in Fig. \ref{fig:shark_fin}, this being the well-known \emph{shark fin diagram}. The shaded region in the diagram corresponds to classical black hole solutions, while the unshaded areas represent solutions containing naked singularities. Compared to AdS, where solutions exist for all values of $M$, in dS it is impossible to have arbitrarily large values for the mass and charge. Generally speaking, each horizon possesses its own temperature, implying the absence of thermal equilibrium \cite{Banihashemi:2022jys}. In addition, for dS the notion of mass is ambiguous due to subtleties involved with asymptotic regions where a conserved mass can be defined.

Each of the three horizons $r_{\text{h}} = \{r_{-}, r_{+}, r_{c}\}$ has an associated temperature $T_{h}$, which is given by
\begin{equation}
    T_{\text{h}} = \frac{1}{4\pi}\left|f'(r)\right|_{r=r_{\text{h}}}\,, 
\end{equation}
For later use, we note here that the temperatures associated with the outer and cosmological horizon $r_{+}$ and $r_{c}$ are
\begin{subequations}
\begin{align}
    T_{+} &= \frac{\left(r_c - r_+\right) \left(r_+ - r_-\right) \left(r_c + 2 r_+ + r_- \right)}{4 \pi  \ell_4^2\, r_{+}^2}\,, \label{eq:temperature_outer}\\
    T_{c} &= -\frac{\left(r_c - r_+\right) \left(r_c - r_-\right) \left(2 r_c + r_+ + r_- \right)}{4 \pi  \ell_4^2\, r_{+}^2}\,.\label{eq:temperature_cosmo}
\end{align}
\end{subequations}
In addition, entropy $S_{h}$ and chemical potential $\mu_h$ for each horizon is defined as
\begin{align}
    S_{h} = \pi r_{h}^2\ \quad \text{and} \quad \mu_h=\frac{Q}{r_h}\,.
\end{align}
The above definitions lead to the first law of black hole thermodynamics
\begin{align}\label{eq:1stlaw}
    dM=\pm T_{h} dS_{h}+\mu_h dQ\,,
\end{align}
where positive sign in the first term on the right hand side of equation \eqref{eq:1stlaw} is considered when $r_h=r_+$, and for the other two cases one should consider the negative sign \cite{Gibbons:1977mu}. For the cosmological horizon it implies that the entropy associated with it is reduced by the addition of Killing energy. Numerous thermodynamic interpretations of the minus sign have been proposed in the literature \cite{Spradlin:2001pw, Klemm:2004mb, Anninos:2012qw, Banihashemi:2022htw}.

\noindent
\section{Near-extremal, near-horizon geometry of RN-\MakeLowercase{d}S$_4$ black hole} \label{sec:extremalsec}
Having reviewed the general properties of RN-dS$_4$ black holes, we now turn to their extremal limits. Extremality occurs when two or more horizons \emph{coalesce}, causing the Hawking temperature associated with both horizons to vanish. For the case of RN-dS$_4$ black holes, naturally three different extremal scenarios are possible:
\begin{enumerate}
    \item \textbf{Cold black hole:} The outer horizon and inner horizon coalesce, i.e. $r_{-} = r_{+} \equiv r_0$. 
    \item \textbf{Nariai black hole:} The outer horizon coalesces with the cosmological horizon, i.e. $r_{+} = r_{c} \equiv r_{\text{n}}$, and
    \item \textbf{Ultracold black hole:} When all three horizons coalesce, i.e. $r_{-} = r_{+} = r_{c} \equiv r_{\text{uc}}$.
\end{enumerate}
These three limits correspond to the two edges and the tip of the shark fin diagram given in Fig.~\ref{fig:shark_fin}. 
\begin{figure}[h]
\centering
\includegraphics[scale=0.45]{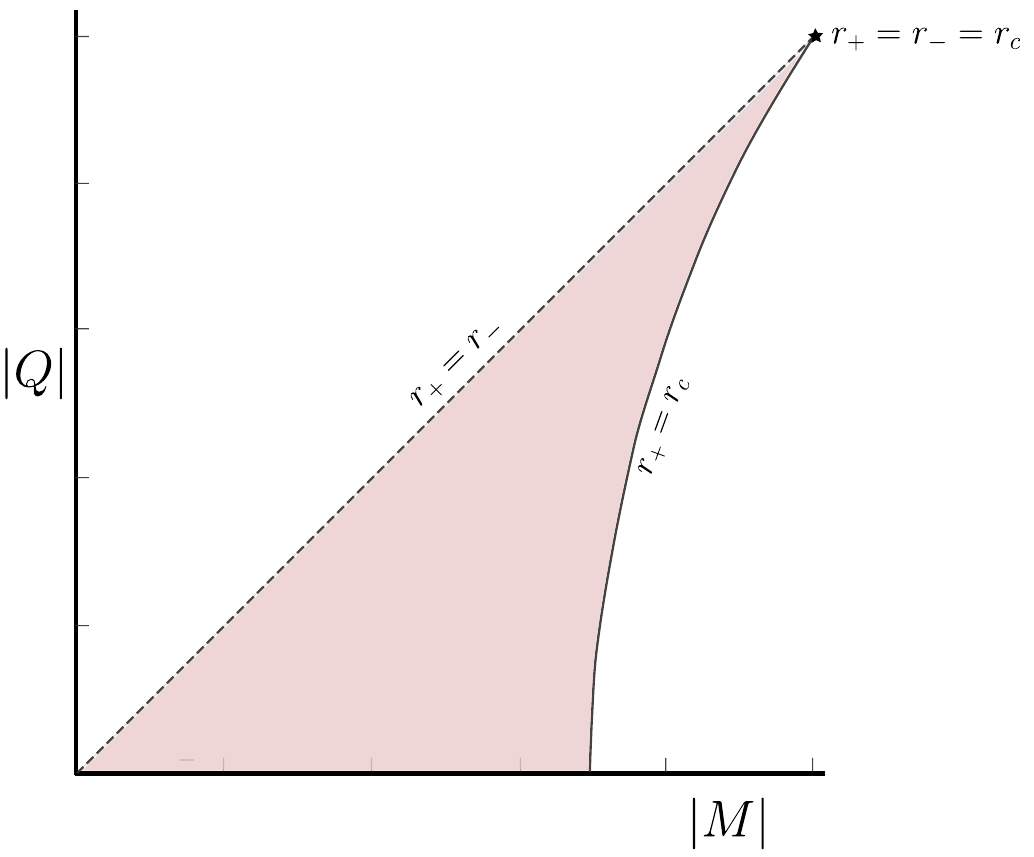}
\caption{\small{The $(M,Q)$ parameter space of classical RN-dS$_4$ black hole solutions. The boundaries of the shaded region correspond to the three extremal limits. The dashed line are the cold black holes, the solid line are the Nariai black holes, and the star where the two lines intersect is the ultracold limit.}}
\label{fig:shark_fin}
\end{figure}

Each extremal case admits a decoupling limit i.e., a clean separation between the near-horizon region and asymptotic region at large radial distances, where the symmetries of the near-horizon geometry of the black hole are enhanced compared to the parent solution. Subsequently, we deal with each of the extremal limits case-by-case and figure out the small temperature correction $0<T \ll 1$ to the near-horizon solution when we move slightly away from extremality. These corrections will eventually lead to the $\log T$ correction in the black hole entropy alluded to in the beginning.

Following the approach of \cite{Iliesiu:2020qvm, Iliesiu:2022onk, Banerjee:2023gll, Kapec:2023ruw, Rakic:2023vhv, Banerjee:2023quv, Maulik:2024dwq}, we write an explicitly temperature-dependent coordinate transformation in each of the three extremal cases, which gives the requisite near-extremal limit while simultaneously zooming in on the near-horizon region also. We emphasize at this point that in all except the ultracold black hole case, we move away from the extremal limit, keeping the electric charge $Q$ fixed. In other words, we describe the black hole in the canonical ensemble; this is simply a convenient choice, since previous works \cite{Iliesiu:2020qvm, Banerjee:2023gll, Kapec:2023ruw, Rakic:2023vhv, Maulik:2024dwq} have demonstrated that the $\log T$ contribution to black hole entropy from the graviton sector happens to be universal, i.e. the choice of the ensemble has no effect on the correction. The ultracold black hole happens to be a single point on the $(M,Q)$ parameter space, and veering away from this limit while staying within the space of black hole solutions inevitably requires us to consider a correction in the electric charge.\vspace{1em}

\subsection{Cold black hole}\vspace{1em}

The cold black hole is obtained when the outer and inner horizons coincide i.e. $r_{-} = r_{+} \equiv r_{0}$. Such a spacetime is represented by a Penrose diagram given in Fig. \ref{fig:cold}. This is the analogous limit that appears in the context of asymptotically AdS and flat black holes. 
\begin{figure}[h]
\centering
\includegraphics[width=0.7\linewidth]{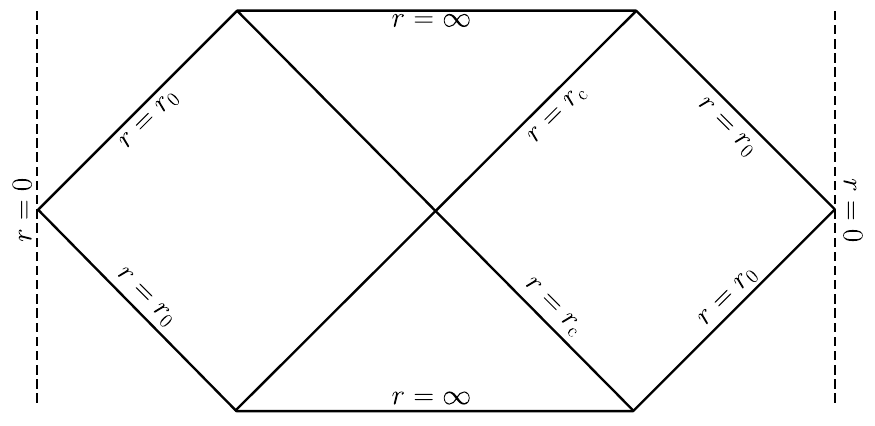}
\caption{Penrose diagram of \emph{cold black hole}, where the outer and inner horizons have merged $r_{+}=r_{-}=r_0$.}
\label{fig:cold}
\end{figure}
Using equation \eqref{eq:temperature_outer}, it is straightforward to show that the inner and outer horizon radii admit the following series expansion when we introduce a small temperature $T_{+}=T$. Namely,
\begin{subequations}
\begin{align}
    r_{+} &= r_{0} + \frac{2\pi r_{0}^2}{1 - \frac{6 r_{0}^2}{\ell_4^2}}\,T + \cO\left(T^2\right)\,, \label{eq:rp_expanded}\\
    r_{-} &= r_{0} - \frac{2\pi r_{0}^2}{1 - \frac{6 r_{0}^2}{\ell_4^2}}\,T + \cO\left(T^2\right)\,.
\end{align}
\end{subequations}
The hierarchy between inner and outer horizon implies $\ell_4^2 > 6r_0^2$. Using equation \eqref{eq:curvature_and_rc} we find that the cosmological horizon also receives correction at $\cO\left(T^2\right)$ given by
\begin{equation}
    r_{c} = -r_{0}+\sqrt{\ell_4^2-2r_{0}^2} - \frac{2 \pi^2 \ell_4^4 r_{0}^3 \left(\ell_4^2\left(3 r_{0}+\sqrt{\ell_4^2-2 r_{0}^2}\right)-2 r_{0}^2\left(5 r_{0}+\sqrt{\ell_4^2- 2 r_{0}^2}\right)\right)}{\left(\ell_4^2-6 r_{0}^2\right)^3 \sqrt{\ell_4^2-2 r_{0}^2}}\,T^2.
\end{equation}
As emphasized before, working in the canonical ensemble fixes the electric charge at the exact extremal limit to 
\begin{equation}
  \left|Q\right|_{\rm{E}} = r_{0} \sqrt{1-\frac{3 r_{0}^2}{\ell_4^2}}\ ,
\end{equation}
while the mass $M$ receives a correction at quadratic order in temperature given by
\begin{equation}
   M_{\rm{NE}} = r_{0}\left(1 -\frac{2 r_{0}^2}{\ell_4^2}\right) + \frac{2 \pi ^2 \ell_4^2 r_{0}^3}{\ell_4^2-6 r_{0}^2}T^2\,.
\end{equation}
We now perform a Wick rotation in the time coordinate in order to Euclideanize the background spacetime. Towards this end, we define a transformation $\{t, r, \theta, \phi\} \rightarrow \{\tau, y, \theta, \phi\}$ on the RN-dS$_4$ solution
\begin{equation}
    t = -\frac{i\tau}{2\pi\,T}\,, \qquad 
    r = r_{+} + \frac{2\pi r_{0}^2}{1-\frac{6 r_{0}^2}{\ell_4^2}}\left(y-1\right)T\,,
\end{equation}
where $r_{+}$ is given by \eqref{eq:rp_expanded}. In the new coordinate system $\tau$ can be interpreted as Euclidean time while $y \in \big[1, \infty\big)$, with $y=1$ surface being the outer horizon. Expanding the metric and the gauge field upto $\cO(T)$ gives 
\begin{equation}
    \begin{split}
        ds^2 &= \bar{g}_{\mu\nu} dx^{\mu} dx^{\nu} + T\,\delta g_{\mu\nu} dx^{\mu} dx^{\nu}\,,\\
        A &= \bar{A} + T\, \delta A\,,
    \end{split}
\end{equation}
where
\begin{subequations}
\begin{align}
    \bar{g}_{\mu\nu} dx^{\mu} dx^{\nu} &= \frac{r_{0}^2}{1 - \frac{6 r_{0}^2}{\ell_4^2}}\left(\left(y^2-1\right)d\tau^2 + \frac{dy^2}{y^2-1} \right) + r_{0}^2\, d\Omega_2^2\,, \label{eq:NHE_metric_C}\\
    \delta g_{\mu\nu} dx^{\mu} dx^{\nu} &= \frac{4\pi r_{0}^3}{1-\frac{6 r_{0}^2}{\ell_4^2}}\left( \frac{\left(1-\frac{4 r_{0}^2}{\ell_4^2}\right)\left(y+2\right)}{\left(1-\frac{6 r_{0}^2}{\ell_4^2}\right)^2} \left(-\left(y-1\right)^2 d\tau^2 + \frac{dy^2}{\left(y+1\right)^2} \right) + y\, d\Omega_2^2 \right)\,,\label{eq:NHNE_metric_C}\\
    \bar{A} &= \frac{i r_{0}\sqrt{1 - \frac{3 r_{0}^2}{\ell_4^2}}}{1-\frac{6r_{0}^2}{\ell_4^2}}\left(y-1\right)d\tau\,, \label{eq:NHE_gaugefield_C}\\
    \delta A &= \frac{2\pi i r_{0}^2\sqrt{1-\frac{3r_{0}^2}{\ell_4^2}}}{\left(1-\frac{6r_{0}^2}{\ell_4^2}\right)^2}\left(y^2-1\right)d\tau\,.\label{eq:NHNE_gaugefield_C}
\end{align}
\end{subequations}
For the sake of completeness, we also note down the $\cO(T)$ correction to the field strength
\begin{equation} \label{eq:NHNE_FS_C}
    F = -\frac{i r_{0} \sqrt{1-\frac{3r_{0}^2}{\ell_4^2}} \left(1-\frac{6r_{0}^2}{\ell_4^2}-4\pi T r_{0}y \right)}{\left(1-\frac{6r_{0}^2}{\ell_4^2}\right)^2}d\tau \wedge dy\,.
\end{equation}
From \eqref{eq:NHE_metric_C} it is evident that the topology of the near-horizon extremal geometry is $\mathrm{AdS}_2 \times \mathrm{S}^2$, with the AdS$_2$ radius being given by
\begin{equation}
    \ell_{\text{AdS}}^2 = \frac{r_{0}^2}{1-\frac{6r_{0}^2}{\ell_4^2}}\,.
\end{equation}
Requiring $\ell_{\text{AdS}}^2>0$ leads us again to the constraint $\ell_{4}^2 > 6 r_{0}^2$, further implying positivity of electric charge and the mass at extremality\footnote{The latter can also be concluded by demanding positivity of $r_{c}$ at extremality from \eqref{eq:curvature_and_rc}.}. Therefore, there occurs a symmetry enhancement in the near-horizon region of the extremal cold black hole spacetime  to SL$\left(2, \mathbb{R}\right) \times$ SO$\left(3\right)$ from the initial U$(1) \times$ SO$(3)$ isometry associated with a generic RN-dS$_4$ black hole. The presence of small temperature correction gives a very large, though finite, AdS$_2$ throat.
Finally, the $\cO(T)$ correction to the Bekenstein-Hawking entropy of the outer horizon resulting from the near-extremal expansion is
\begin{equation}
    S_{+} = \pi r_{0}^2 + \frac{4\pi^2r_{0}^3}{1-\frac{6r_{0}^2}{\ell_4^2}}T\,.
\end{equation}

\subsection{Nariai black hole}
\label{sec:NariaiNHNE}\vspace{1em}
The Nariai limit corresponds to the case when the outer horizon and the cosmological horizon coalesce, i.e. $r_{+} = r_{c} \equiv r_{\mathrm{n}}$. Fig. \ref{fig:Nariai} represents the corresponding Penrose diagram of the Nariai black hole (at exact extremality).

\begin{figure}[h]
\centering
\includegraphics[width=0.5\linewidth]{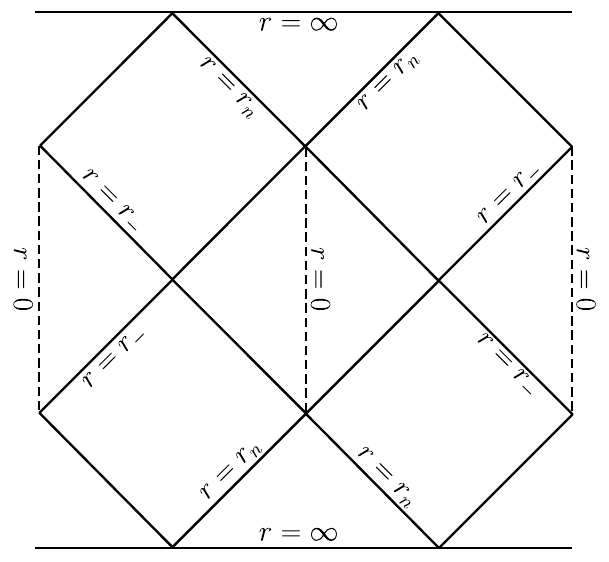}
\caption{Penrose diagram of charged Nariai black hole}
\label{fig:Nariai}
\end{figure}

The temperature associated to both the outer as well as cosmological horizon vanish in this limit (follows from \eqref{eq:temperature_outer} and \eqref{eq:temperature_cosmo}). It is important to note that, in contrast to AdS or flat cases, the dS Schwarzschild black hole allows us to observe the Nariai limit due to the presence of two horizons \cite{Cho:2007we}. Working in the canonical ensemble (keeping the charge fixed) one can perform a series expansion near extremality. 

An analogous computation to the cold black hole case, leads us to the following series for $r_{+}$ and $r_{c}$
\begin{equation}
    r_{+} = r_{\mathrm{n}} - \frac{2\pi\ell_4^2 r_{\mathrm{n}}^2}{6 r_{\mathrm{n}}^2 - \ell_4^2}\,T + \cO\left(T^2\right)\,,\quad 
    r_{c} = r_{\mathrm{n}} + \frac{2\pi\ell_4^2 r_{\mathrm{n}}^2}{6 r_{\mathrm{n}}^2 - \ell_4^2}\,T + \cO\left(T^2\right)\,,
\end{equation}
where the temperature associated with $r_{+}$, $T_{+}=T$ is introduced as a small deviation\footnote{We should contrast this with a very similar recent study performed in \cite{Blacker:2025zca}. In this study, the observer is in the Milne (or inflationary) patch thus increasing $T_c$ by a small deviation. However our observer initially is in the static patch which is bounded by two horizons \cite{Maldacena:2019cbz}, one outer horizon at $r_{+}$ and another cosmological horizon at $r_c$.}. The temperature correction to the inner horizon can be obtained using the two above expansions along with equation \eqref{eq:curvature_and_rc}, giving
\begin{equation}
\begin{split}
    &r_{-} =\\& -r_{\mathrm{n}} + \sqrt{\ell_4^2 - 2 r_{\mathrm{n}}^2} + \frac{2 \pi ^2 \ell_4^4 r_{\mathrm{n}}^3 \left(\ell_4^2 \left(\sqrt{\ell_4^2-2 r_{\mathrm{n}}^2}+3 r_{\mathrm{n}}\right)-2 r_{\mathrm{n}}^2 \left(\sqrt{\ell_4^2-2 r_{\mathrm{n}}^2}+5 r_{\mathrm{n}}\right)\right)}{\left(6 r_{\mathrm{n}}^2 - \ell_4^2\right)^3 \sqrt{\ell_4^2-2 r_{\mathrm{n}}^2}}\,T^2\,.
\end{split}
\end{equation}
Here we also consider the canonical ensemble keeping the charge fixed at
\begin{align}
     |Q|_E &= \rn\sqrt{1 - \frac{3\rn^2}{\ell_4^2}}\,.
\end{align}
The small temperature corrections to the Bekestein-Hawking entropy of the outer and cosmological horizons near the Nariai limit are given by
\begin{align}
\label{eq:Nariaicharge}
    M|_{\rm{NE}} &= \rn\left(1-\frac{2 \rn^2}{\ell_4^2}\right) - \frac{2 \pi ^2 \ell_4^2 \rn^3}{6 \rn^2-\ell_4^2}\,T^2,\\
\label{eq:NariaiCosmoEntropy}
    S_{+} &= \pi\rn^2 - \frac{4\pi^2\rn^3}{\frac{6\rn^2}{\ell_4^2}-1}\,T\,  ,\qquad
    S_{c} = \pi\rn^2 + \frac{4\pi^2\rn^3}{\frac{6\rn^2}{\ell_4^2}-1}\,T\,.
\end{align}
Imposing positivity of $Q^2$ along with the relative hierarchy of outer and cosmological horizons i.e. $r_c > r_{+}$ puts an upper as well as lower bound of $r_{\rm n}$, namely, $\frac{\ell_4^2}{6}<\rn^2<\frac{\ell_4^2}{3}$. This is distinct from the previous case of cold black hole where there was only an upper bound on the extremal horizon radius. We now introduce the following coordinate transformations to Euclideanize our background spacetime
\begin{align}
    t = -\frac{i\tau}{2\pi T}\,, \qquad r = r_{+} - \frac{2\pi \ell_4^2 r_{\mathrm{n}}^2}{6 r_{\mathrm{n}}^2 - \ell_4^2}(y-1)\,T\,=\rn - \frac{2\pi \ell_4^2 r_{\mathrm{n}}^2}{6 r_{\mathrm{n}}^2 - \ell_4^2}y\,T\, ,
\end{align}
and expand the transformed metric and gauge field upto linear order in temperature.  Here $y \in \big(-1,1\big)$, with $y=1$ corresponding to exact extremality. The transformation of the radial variable is configured in a way such that the Nariai radius $r_{\rm n}$ is approached from the interior in accordance with \cite{Castro:2022cuo}. Once again we arrive at
\begin{equation}
    \begin{split}
        ds^2 &= \bar{g}_{\mu\nu} dx^{\mu} dx^{\nu} + T\,\delta g_{\mu\nu} dx^{\mu} dx^{\nu},\\
        A &= \bar{A} + T\, \delta A\,,
    \end{split}
\end{equation}
where the zeroth and first order terms of the metric have the following expressions
\begin{align}
    \bar{g}_{\mu\nu} dx^{\mu} dx^{\nu} &= \frac{r_{\mathrm{n}}^2}{\frac{6 r_{\mathrm{n}}^2}{\ell_4^2}-1}\left(\left(1-y^2\right) d\tau^2 + \frac{dy^2}{1-y^2}\right) + r_{\mathrm{n}}^2\, d\Omega_2^2\,, \label{eq:NHE_metric_N}
    \end{align}
    \begin{align}
        \delta g_{\mu\nu} dx^{\mu} dx^{\nu} = \frac{4\pi r_{\mathrm{n}}^3}{\frac{6 r_{\mathrm{n}}^2}{\ell_4^2}-1}\left( \frac{\left(1-\frac{4 r_{\mathrm{n}}^2}{\ell_4^2}\right)}{\left(\frac{6 r_{\mathrm{n}}^2}{\ell_4^2}-1\right)^2} \left(-y\left(y^2-1\right) d\tau^2 + \frac{y dy^2}{\left(y^2-1 \right)}\right) + y\, d\Omega_2^2 \right)\, ,  \label{eq:NHNE_metric_N}
    \end{align}
The gauge field background solution and the first order correction are given by
    \begin{align}
    \bar{A} &= -i|Q|_{\rm{E}}\frac{y}{\frac{6r_{\mathrm{n}}^2}{\ell_4^2}-1}d\tau=\frac{i r_{\mathrm{n}}\sqrt{1 - \frac{3 r_{\mathrm{n}}^2}{\ell_4^2}}}{\frac{6r_{\mathrm{n}}^2}{\ell_4^2}-1}(1-y)~d\tau\,, \label{eq:NHE_gaugefield_N}\\
    \delta A &= \frac{2\pi i r_{\mathrm{n}}^2\sqrt{1-\frac{3r_{\mathrm{n}}^2}{\ell_4^2}}}{\left(\frac{6r_{\mathrm{n}}^2}{\ell_4^2}-1\right)^2}(1-y^2) d\tau\,. \label{eq:NHNE_gaugefield_N}
\end{align}
The corresponding field strength (up to $\cO(T)$ terms) turns out to be
\begin{equation} \label{eq:NHNE_FS_N}
    F = \frac{i \rn \sqrt{1 - \frac{3 \rn^2}{\ell_4^2}} \left(\frac{6 \rn^2}{\ell_4^2} - 1 + 4 \pi  \rn\,y\, T \right)}{\left(\frac{6 \rn^2}{\ell_4^2}-1\right)^2} d\tau \wedge dy\,.
\end{equation}
It follows from \eqref{eq:NHE_metric_N} that the near-horizon metric at the Nariai limit \eqref{eq:NHE_metric_N} has the topology of dS$_2 \times$ S$^2$, with the curvature scale of the two-dimensional de Sitter space being set by
\begin{equation} \label{eq:ds_curvature}
    \ell_{\text{dS}}^2 = \frac{\rn^2}{\frac{6\rn^2}{\ell_4^2}-1}\,>0\,,
\end{equation}
which is positive definite due to the bound on $r_{\rm n}$. {A few crucial comments regarding the Nariai case are in order. Note that in \eqref{eq:NHE_metric_N} the dS$_2$ metric is in static coordinates for Lorentzian time and range of $y$ implies a compact manifold. We will subsequently see that in order to obtain the quantum correction to the thermodynamic entropy for the near-extremal Nariai black hole, we need to complexify and extend the contour of the coordinate $y$ from the cosmological horizon to complex infinity. 
`Gluing' of an Euclidean and Lorentzian dS$_2$ helps us to understand the quantum correction of Hartle-Hawking like wave function.
We expand on this in the subsequent section and the leading behaviour of the Hartle-Hawking like wave function at the Nariai limit is provided in Appendix \ref{appendix:B}.}\vspace{1em}

\subsection{Ultracold black hole} \label{sec:sec3_subsec3}\vspace{1em}
The ultracold limit is characterized by the coinciding of \emph{all three horizons} i.e. $r_{-} = r_{+} = r_{c} = \ruc=\ell_4/\sqrt{6}$. 
The ultracold black hole corresponds to a single point located at the tip of the \emph{shark-fin diagram} (Fig. \ref{fig:shark_fin}). The black hole has a Penrose diagram given in Fig.~\ref{fig:ultracold}.
\begin{figure}[H]
\centering
\includegraphics[width=0.32\linewidth]{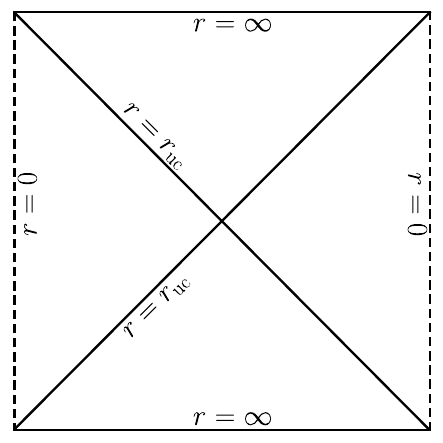}
\caption{Penrose diagram of ultracold black hole}
\label{fig:ultracold}
\end{figure}
Let us remind the reader that the space of black hole solutions is the area inside the shark-fin diagram bounded by the lines of extremality (for cold and Nariai black holes) and the horizontal axis. It is impossible to study deviations from the ultracold limit while keeping either the electric charge or the mass fixed. Thus, one may consider to work in the grand canonical ensemble where both $Q$ and $M$ are allowed to vary such that one stays inside the shaded area of the shark-fin diagram. Schematically, the deviations in all three cases are denoted by the arrowheads in Fig. \ref{fig:deviations_from_extremality}.\vspace{1em}
\begin{figure}[H]
\centering
\includegraphics[scale=0.45]{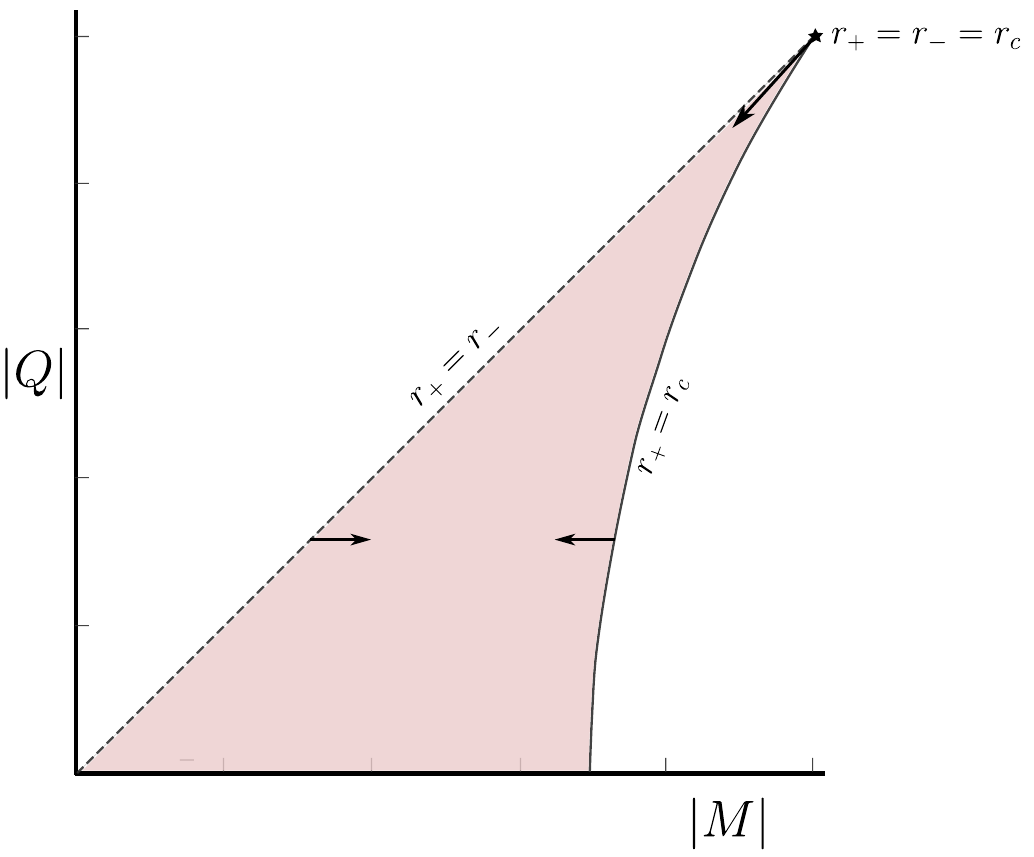}
\caption{\small{Moving away from the three extremal limits. In the cold and Nariai case, one may choose to keep $\left|Q\right|$ constant and change the mass $M$ with temperature, and still remain within the space of black hole solutions. In the ultracold limit, however, both $|Q|$ and $M$ must change.}}
\label{fig:deviations_from_extremality}
\end{figure}
As in previous examples, we perform a \emph{small temperature} expansion of all parameters about their extremal ultracold values. However, unlike the previous two cases, here the expansion needs to be done in $\sqrt{T}$ where $T \equiv T_{+}$ is the temperature associated with the outer horizon. Such a form of the expansion is in alignment with previous work \cite{Castro:2022cuo} and a detailed explanation for such an expansion is provided in Appendix \ref{appendix:A}.

We start with the following expansion for the three \emph{almost} coinciding horizons
\begin{subequations}
\begin{align}
    r_{-} &= \frac{\ell_4}{\sqrt{6}} + \beta_{1}\sqrt{T} +  \cO\left(T \right)\,,\\
    r_{+} &= \frac{\ell_4}{\sqrt{6}} + \alpha_{1}\sqrt{T} + \cO\left(T \right)\,,\\
    r_{c} &= \frac{\ell_4}{\sqrt{6}} + \left(-\alpha_{1} - \beta_{1} \right) \sqrt{T} + \cO\left(T\right)\,.
\end{align}
\end{subequations}
The third equation above results from \eqref{eq:curvature_and_rc}. 
Note here, unlike the previous two cases, the ultracold black hole solution is determined by a single scale $\ell_4$ (or equivalently $r_{\text{uc}}$). Demanding consistency of the above series expansion with \eqref{eq:temperature_outer} along with the demand $r_{+}>r_{-}$ fixes $\beta_1$ in terms of $\alpha_1$
\begin{equation}
    \beta_{1} = \frac{1}{6} \left(-3 \alpha_{1} - \sqrt{81 \alpha_{1}^2 + 6 \sqrt{6}\, \pi \ell_{4}^3}\right).
\end{equation}
In the case of the near cold and near Nariai black holes, the constant $\alpha_1$ was further fixed by demanding the electric charge $Q$ be independent of temperature (since we were in a canonical ensemble). In the present case, we no longer have that freedom. Instead, we demand constancy of the chemical potential at the outer horizon i.e. $\mu_{+} = \frac{Q}{r_{+}}$ stays the same away from extremality at first order in the expansion parameter (in other words, the coefficient of the $\cO(\sqrt{T})$ term in the $\mu_+$ expansion vanished identically). This eventually gives us,
\begin{equation}
    \alpha_{1} = 0\,.
\end{equation}
Note that the choice of keeping $\mu_{+}$ fixed up to $\cO(\sqrt{T})$ is done solely for the purpose of fixing the undetermined constant $\alpha_1$. Interestingly, we will subsequently observe that the leading order correction to entropy of the outer horizon appears at $\cO(T)$. The small temperature expansion eventually takes the following form
\begin{subequations}
    \begin{align}
    \label{eq:ucrminus}
    r_{-} &= r_{\rm uc} - \sqrt{\frac{\pi \ell_{4}^3}{\sqrt{6}}} \sqrt{T} +  \cO\left(T \right)\, ,\\
    r_{+} &= r_{\rm uc} + \cO\left(T \right)\, ,\\
    \label{eq:ucrcosmo}
    r_{c} &= r_{\rm uc} + \sqrt{\frac{\pi \ell_{4}^3}{\sqrt{6}}} \sqrt{T} + \cO\left(T\right)\, ,
    \end{align}
\end{subequations}
where $r_{\rm uc}=\frac{\ell_4}{\sqrt{6}}$. Borrowing inspiration from \cite{Castro:2022cuo}, we consider the change of coordinates
\begin{eqnarray}
    t&=& \pm \frac{i\kappa}{T^{3/4}}\tau\,,\\
    r&=& r_{\rm uc}-R_0 \sqrt{T}\pm \frac{T^{3/4}}{\kappa}y\,,
\end{eqnarray}
where $\kappa=\left[ \frac{3r_{\rm uc}^3}{2R_0^3-12\pi R_0 r_{\rm uc}^3}\right]^{1/2}$. Plugging in the above in \eqref{eq:RNdS4_solution} and expanding up to the first subleading order in the metric gives us \begin{equation}
        ds^2 = \bar{g}_{\mu\nu} dx^{\mu} dx^{\nu} + T^{\frac{1}{4}}\,\delta g_{\mu\nu} dx^{\mu} dx^{\nu},
\end{equation}
We have fixed the coefficients appearing in the above transformations in order to obtain Mink$_2$ (or $\mathbb{R}^2$, since we have performed a Wick rotation) as the leading effective 2-dimensional near-horizon extremal metric. 
More explicitly, the zeroth and first order terms are
\begin{subequations}
\begin{align}
    &\bar{g}_{\mu\nu} dx^{\mu} dx^{\nu} = d\tau^2 + dy^2 + r_{\mathrm{uc}}^2\, \left(d\theta^2 + \sin^2\theta d\phi^2\right)\,, \label{eq:NHE_metric_UC}\\
    \begin{split}
    &\delta g_{\mu\nu} dx^{\mu} dx^{\nu} = \frac{\sqrt{6} \left(R_{0}^2 - 2\pi\ruc^3 \right)}{\sqrt{\ruc^{3} \left(R_{0}^3 - 6\pi R_{0} \ruc^3 \right)}} \left(-y\, d\tau^2 + y\, dy^2 \right)\,.
    \end{split} \label{eq:NHNE_metric_UC}
\end{align}
\end{subequations}

The near-horizon extreme spacetime in the ultracold case is a two-dimensional Euclidean plane times a sphere $\left(\mathbb{E}^2\times S^2\right)$. Curiously, we see the gauge field expansion is divergent in the $T\to 0$ limit
\begin{equation}
    A = \frac{i}{2} \left(- \sqrt{\frac{3 R_{0} \ruc}{R_{0}^2 - 6\pi\ruc^3}}\ T^{-\frac{1}{4}} + \frac{\sqrt{2}\ y}{\ruc} - \sqrt{\frac{3 R_{0}^3}{\ruc \left(R_{0}^2 - 6\pi\ruc^3 \right)}}\ T^{\frac{1}{4}} \right)d\tau\ + \cO\left(\sqrt{T}\right) .
\end{equation}
although the field strength remains finite
\begin{equation}
    F = - \frac{i}{\sqrt{2}\,\ruc} d\tau \wedge dy + \mathcal{O}\left(\sqrt{T}\right)\,.
\end{equation}
Since the divergent part in the gauge field is constant, it can be gauged away for any small temperature. The thermodynamic quantities in the near-extremal ultracold limit are
\begin{align}
    Q &= \frac{\ell_4}{2\sqrt{3}} - \frac{\sqrt{2} \pi \ell^{2}_{4}}{3}\,T\,,\\
    M &= \sqrt{\frac{2}{3}}\frac{\ell_4}{3} - \frac{\pi \ell_{4}^{2}}{3}\,T\,,\\
    S_{+} &= \frac{\pi\ell_4^2}{6} +  \cO\left(T\right)\,.
\end{align}
One should contrast the above expressions with their cold and Nariai counterparts. Here, the expansion parameter being $\sqrt{T}$, the leading order corrections in $Q$, $M$ as well as $S_+$ appears at $\cO(T)$ i.e. quadratic in the expansion parameter, although the expansions associated with the inner ($r_{-}$) and cosmological ($r_c$) horizons had $\cO(\sqrt{T})$ terms.

If one wants to describe the region between the outer horizon ($r_{+}$) and the cosmological horizon ($r_c$) in terms of the expansions as defined in \eqref{eq:ucrminus}-\eqref{eq:ucrcosmo}, we see that the new radial coordinate $y$ is restricted to the interval\footnote{One can further rescale the radial coordinate as $y_1=y\frac{T^{1/4}}{R_0\kappa}$, which changes its range to
\begin{equation}
\begin{aligned}
   1< y \frac{T^{1/4}}{R_0\kappa}< 1+\frac{1}{R_0}\sqrt{\frac{\pi \ell_4^3}{\sqrt{6}}}\quad \Rightarrow\quad 1< y_1 < 1+\frac{1}{R_0}\sqrt{\frac{\pi \ell_4^3}{\sqrt{6}}}
\end{aligned}
\end{equation}. This however changes the canonical form of the $\mathbb{R}^2$ metric into $d\tau^2+\alpha^2 dy_1^2$, for some constant $\alpha$.}
\begin{equation}
    \frac{R_0 \kappa}{T^{1/4}}< y< \frac{R_0 \kappa}{T^{1/4}}+\frac{\kappa}{T^{1/4}}\sqrt{\frac{\pi \ell_4^3}{\sqrt{6}}}\,.
\end{equation}


\section{The extremal zero modes of RN-\MakeLowercase{d}S$_4$ black hole} \label{sec:sec3}

We now illustrate how the $\log T$ correction to the extremal thermodynamic entropy arises from the zero modes of the kinetic operator within the path integral framework. This analysis incorporates the small-temperature corrections obtained in the previous section. Using the saddle point approximation, we can express the one-loop contribution to the generating functional as a Gaussian integral over the linearized metric perturbations around the background $\bar{g}$
\begin{equation}\label{1loopPI}
   Z\approx \exp\left(-I[\bar{g}]\right)\int [Dh]\exp \left[-\int d^4x \sqrt{\bar{g}}\,h_{\alpha\beta}^*\Delta^{\alpha\beta\mu\nu}[\bar{g}]h_{\mu\nu}\right] \; .
\end{equation}
Here $\Delta^{\alpha\beta\mu\nu}$ is the extremal Lichnerowicz operator and can be written as the following combination \cite{Kapec:2023ruw, Rakic:2023vhv, Maulik:2024dwq}
 \begin{equation}\label{eq:Lichnerowicz_op}
     \Delta^{\alpha\beta\mu\nu}= \Delta^{\alpha\beta\mu\nu}_{\rm Ein}+\Delta^{\alpha\beta\mu\nu}_{\Lambda}+\Delta^{\alpha\beta\mu\nu}_{\rm Max}\,, 
 \end{equation}
 with
 \begin{equation}
    \begin{aligned}
       &\Delta^{\alpha\beta\mu\nu}_{\rm Ein}=\left(\frac{1}{2} \bar{g}^{\alpha \mu} \bar{g}^{\beta \nu} \bar{\square} -\frac{1}{4} \bar{g}^{\alpha \beta} \bar{g}^{\mu \nu} \bar{\square} + \bar{R}^{\alpha \mu \beta \nu} +\bar{R}^{\alpha \mu} \bar{g}^{\beta \nu} - \bar{R}^{\alpha \beta} \bar{g}^{\mu \nu}  -\frac{1}{2} \bar{R} \bar{g}^{\alpha \mu} \bar{g}^{\beta \nu} + \frac{1}{4} \bar{R} \bar{g}^{\alpha \beta} \bar{g}^{\mu \nu} \right)\,,\\&\Delta^{\alpha\beta\mu\nu}_{\Lambda}=\left(\bar{g}^{\alpha \mu} \bar{g}^{\beta \nu} -\frac{1}{2} \bar{g}^{\alpha \beta} \bar{g}^{\mu \nu}\right)\Lambda\,,\\&\Delta^{\alpha\beta\mu\nu}_{\rm Max}=\frac{1}{8}F^2(2\bar{g}^{\alpha\mu}\bar{g}^{\beta\nu}-\bar{g}^{\alpha\beta}\bar{g}^{\mu\nu})-F^{\alpha\mu}F^{\beta\nu}-2F^{\alpha\gamma}F^{\mu}{}_{\gamma}\bar{g}^{\beta\nu}+F^{\alpha\gamma}F^{\beta}{}_{\gamma}\bar{g}^{\mu\nu}\,.
    \end{aligned}
\end{equation}
In the above $\bar{\square}$ represents the scalar Laplacian over the background metric. In the following, we will study a set of normalizable tensor zero modes supported by the operator~$\Delta$ for both the cold and Nariai limits. These zero modes are not properly accounted for by the heat kernel and need to be treated separately. They in general arise from large diffeomorphisms generated by non-normalizable vector fields that remain unfixed by the harmonic gauge imposed on the metric perturbations.\vspace{1em}

\subsection{Cold black hole}\vspace{1em}

In the cold black hole case, the near-horizon extremal geometry is~AdS$_2\times$ S$^2$, as given by \eqref{eq:NHE_metric_C}. In this case, the following set of (normalizable) tensor zero modes of the Lichnerowicz operator, where the fluctuation involves only the AdS$_2$ directions, is well-known \cite{Sen:2012kpz, Banerjee:2023gll, Kapec:2023ruw, Rakic:2023vhv} and given by,
\begin{equation}\label{eq:zeromode_C}
    h_{\mu\nu}dx^{\mu}dx^{\nu} = \frac{\sqrt{|n|\left(n^2-1\right)}}{2\sqrt{2}\pi\sqrt{1-\frac{6r_0^2}{\ell_4^2}}} \left(\frac{y-1}{y+1}\right)^{\frac{|n|}{2}} e^{in\tau} \left(-d\tau^2 + \frac{2\, i\, d\tau dy}{y^2-1} + \frac{dy^2}{\left(y^2-1\right)^2}\right)\,,
\end{equation}
where $\left|n\right| \geq 2$. The modes $n=0, \pm 1$ need to be excluded or equivalently one need to take a quotient of the path integral measure by SL$\left(2, \mathbb{R}\right)$ group since these perturbations correspond to diffeomorphisms
generated by the isometries of background metric $\bar{g}_{\mu\nu}$.

It can be shown that these zero modes arise from large diffeomorphisms of the AdS$_2$ part of the extreme near-horizon geometry. The diffeomorphisms are driven by the vector field
\begin{equation}
    \xi\left(\tau,y\right) = e^{in\tau}\left(f_{1}\left(y\right)\partial_{y} + f_{2}\left(y\right)\partial_{\tau}\right)\,,
\end{equation}
where
\begin{align}
\label{eq:large_gauge_VF_C_1}
    f_{1}\left(y\right) = \left(\frac{y-1}{y+1}\right)^{\frac{\left|n\right|}{2}} \frac{|n|+y}{2\left(n^2-1\right)}\,,  \qquad 
    f_2(y) = \frac{i\,f'_{1}(y)}{|n|}\,.
\end{align}
The vector field itself is non-normalizable and acts nontrivially on the asymptotic boundary while the modes \eqref{eq:zeromode_C} obtained from $h_{\mu\nu} = \mathcal{L}_{\xi}\bar{g}_{\mu\nu}$ are normalizable.
\vspace{1em}
\subsection{Nariai black hole}\vspace{1em}

As discussed in the \ref{sec:NariaiNHNE}, the near-horizon spacetime geometry in the near-Nariai limit \eqref{eq:NHE_metric_N} has the product topology of dS$_2\times $S$^2$. In \eqref{eq:NHE_metric_N} the radial coordinate $y \in (-1,+1)$ while the Euclidean time coordinate $\tau \in [0,2\pi]$. Naively, it seems that such a compact space which is isomorphic to a 2-sphere\footnote{This can be seen using the transformation $y=\cos \theta$ in \eqref{eq:NHE_metric_N}. } does not admit tensorial zero modes resulting from non-normalizable diffeomorphism generators. In contrast to the AdS case with disk topology, the path integral in dS$_2$ scenario involves spacetime metrics with different signatures. Two distinct contours-- dubbed the Hartle-Hawking contour and the Maldacena contour have been proposed to calculate the no-boundary wavefunction, each reflecting different choices of spacetime evolution. We will consider the conventional Hartle-Hawking contour (depicted by brown line in Fig. \ref{fig:eefig}), which begins with a Euclidean dS space with S$^2$ topology, from the north pole at $y=1$ to equator $y=0$. This Euclidean dS space is then connected at its equator to a Minkowski dS space (obtained following an analytic continuation).

\begin{figure}
   	\centering
   	\begin{tikzpicture}
   	\draw[->,ultra thick] (-5,0)--(5,0) node[right]{};
   	\draw[->,ultra thick] (0,-5)--(0,5) node[above]{};
   	\draw[snake, color=red, thick] (1,0) -- (-5,0);
   	\draw [ color=brown, thick, middlearrow=latex] (1,0.1)
   	{[rounded corners]--(0.2,0.1)--(0.2,5)};
   	\draw [ color=green, thick, 3multiarrow=latex] (1,0.1)
   	{[rounded corners]--(5,0.2)arc(0:90:4.6cm)};
   	\draw [ color=brown, thick, middlearrow=latex] (1,-0.1)
   	{[rounded corners]--(0.2,-0.1)--(0.2,-5)};
   	\draw [ color=green, thick, 3multiarrow=latex] (1,-0.1)
   	{[rounded corners]--(5,-0.2) arc(0:-90:4.6cm)};
   	\draw node[ anchor=south west] at	(3,5) {$y$};
   	\draw (3.05,5.3) -- (3.05,5)--(3.4,5);
   	\fill[black] (1,0) circle (0.1cm);
   	\fill[black] (-1,0) circle (0.1cm);
   	\draw node[ anchor=north ] at	(1,0) {$1$};
   	\draw node[ anchor=north ] at	(-1,0) {$-1$};	
   	\draw node[ anchor=south ] at	(1,0) {$P$};
    \draw node[ anchor=south ] at	(0.5,0) {\small{$\Sigma_1$}};
   	\draw node[ anchor=west ] at	(0.2,5) {$i\infty$};
   	\draw node[ anchor=south east ] at	(0,0) {$O$};
   	\draw node[ anchor=west ] at	(0.2,-5) {$-i\infty$};
   	\draw node[ anchor=west ] at	(5,0) {$\infty$};
    \draw node[ anchor=west ] at	(0.3,2.5) {$\Sigma_2$};
   	\end{tikzpicture}
   	\caption{\small{Analytic continuation in the complex $y$-plane; brown line: Hartle-Hawking contour, green line: Maldacena contour; While $\Re(y)<1$ the geometry in ($y,\tau$) plane is S$^2$ and when $\Re(y)>1$ it is -AdS$_2$}}
   	\label{fig:eefig}
\end{figure}
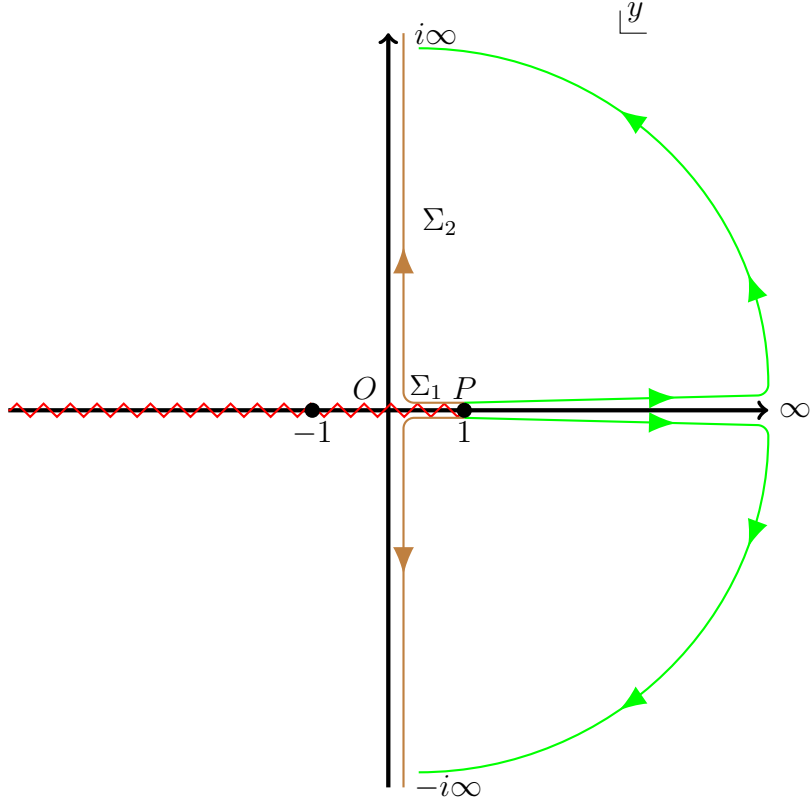

First we focus on the part of the brown contour from the point $P$ to imaginary $y$-axis (close to the point $O$). Over this part of the contour, the metric is that of Euclidean dS$_2$. To be mathematically precise, along this segment, the coordinate $y$ can be parametrized as $y=u+i\delta$, where $\delta>0$ is an infinitesimal constant while $u,\delta \in \mathbb{R}_{+}$. In all subsequent computations, one must take the \emph{constant parameter} $\delta \to 0$. $\delta$ essentially acts as a regulator so that contour does not exactly coincide with the branch cut from $P$ to $O$ (the origin).

In the above parametrization, the background metric takes the form
\begin{equation}
    \bar{g}_{\mu \nu}dx^{\mu}dx^{\nu}=\ell_{\rm dS}^2 \left[(1-u^2)d\tau^2 +\frac{du^2}{1-u^2} \right] +r_{\rm n}^2 d\Omega_2^2 + \cdots\,,
\end{equation}
whose tensor zero modes are given by \footnote{These zero modes are not generated by any non-normalizable vector fields.}
\begin{align}
\left.{h_{\mu\nu}^{(n)}}dx^{\mu}dx^{\nu}\right|_{\Sigma_1} &=
   2 {a}_n e^{in \tau}\frac{\left(n^2-1\right)}{(1-u^2)}  \left(\frac{u-1}{u+1}\right)^{\frac{\left| n\right| }{2}}\left(n (1-u^2)d\tau^2+2i|n|d\tau du+\frac{n}{1-u^2}du^2\right)\nonumber\\
   &\hspace{8cm}+ \cdots\,\,,
\end{align}
where the ellipsis denote terms of $\cO(\delta)$ which eventually vanish under the limit $\delta \to 0$. Recall that the inner product between the tensor modes $h^{(m)}_{\mu \nu}$ is defined as
\begin{equation}
\label{eq:innerproddefn}
    \langle h^{(m)}, h^{(n)} \rangle = \int_{\mathcal{C}}d^4x\ \sqrt{\bar{g}}\bar{g}^{\mu \rho}\bar{g}^{\nu \sigma} {h^{(m)}_{\mu \nu}}^{*} h^{(n)}_{\rho \sigma}=\delta_{m,n}\,,
\end{equation}
where the $*$ denotes the usual complex conjugate. From the above definition, it follows that over the part of the contour $\Sigma_1$, running parallel to the real $y$ axis from the point $P$ to $O$, we get
\begin{equation}
\label{eq:sigma1contrib}
\begin{aligned}
\left.\langle h^{(m)}, h^{(n)} \rangle \right|_{\Sigma_1} &= \frac{8a_m^* a_n r_{\rm n}^2}{\ell_{\rm dS}^2}(m^2-1)(n^2-1)(mn+ |mn|)\,\times\,\\
&\int d^2\Omega_2\int_{\Sigma_1}d\tau\ e^{-i(m-n)\tau} \int_{1}^{0}du\  \left( \frac{u-1}{u+1}\right)^{\frac{1}{2}(|m|+|n|)}\frac{1}{(u^2-1)^2}\\
&=|a_m|^2\frac{128\pi^2 r_{\rm n}^2}{\ell_{\rm dS}^2}m^2(m^2-1)^2\delta_{m,n} \int_{1}^0 du\ \left( \frac{u-1}{u+1}\right)^{|m|}\frac{1}{(1-u^2)^2}\\
&=\delta_{m,n}\frac{32\pi^2 r_{\rm n}^2}{\ell_{\rm dS}^2}|a_m|^2(-1)^{|m|}|m|(m^2-1) (1-2m^2)\,.\\
\end{aligned}
\end{equation}
Now, in order to go over to the dS$_2$ with (1,1) signature, which is glued to the (2,0) dS$_2$ at its equatorial line, we perform an analytic continuation of the radial coordinate. Performing a transformation $y \to iy\,, \,\tau \to \tau\,,$ we see that the analytically continued background metric (follows from \eqref{eq:NHE_metric_N}) of (1,1) signature is given by
\begin{equation}
   \bar{g}_{\mu \nu}dx^{\mu}dx^{\nu}=\ell_{\rm dS}^2 \left[(1+y^2)d\tau^2 -\frac{dy^2}{1+y^2} \right]+r_{\rm n}^2 d\Omega_2^2\,.
\end{equation}
Due to the analytic continuation, now $y$ becomes timelike in the new range along the entire positive imaginary $y$-axis from $i\delta$ to $i\infty$\footnote{At $y=i\infty$ the induced metric will be $ds^2_{\rm induced}=\ell_{\rm dS}^2 (1+y_b^2)d\tau^2 +r_{\rm n}^2 d\Omega_2^2$, which is $S^1\times S^2$ spatial topology with $\ell_{\rm dS} \sqrt{1+y_b^2}$ being the radius of the circle and $y_b$ being the cut off introduced.}. Like before, following \cite{Moitra:2021uiv}, solving for the tensorial zero modes of the Lichnerowicz operator on a dS patch with (1,1) signature gives the solution \footnote{These zero modes are related to AdS$_2$ zero modes under $y\rightarrow -i y$ \cite{Moitra:2021uiv}.} 

\begin{equation}\label{eq:zeromode_N}
\left.h_{\mu\nu}^{(n)}dx^{\mu}dx^{\nu}\right|_{\Sigma_2} = 2a_ne^{in\tau}\frac{n^2-1}{y^2+1}\left(\frac{y-i}{y+i}\right)^{\frac{|n|}{2}}\left(n(y^2+1)d\tau^2 + 2|n|d\tau dy + \frac{n}{y^2+1}dy^2 \right)\,.   
\end{equation}
Note that along the contour $\Sigma_2$, the new $y$ cordinate is \emph{purely imaginary} and hence satisfies $y^{*}=-y$. Therefore, we have,
\begin{align}
\left.{h_{\mu\nu}^{(n)}}^{*}dx^{\mu}dx^{\nu}\right|_{\Sigma_2} &=
   2 {a}_n^{*} e^{-in \tau}\frac{n^2-1}{y^2+1}  \left(\frac{y-i}{y+i}\right)^{\frac{|n|}{2}}\left(n (y^2+1)d\tau^2-2|n|d\tau dy+\frac{n}{y^2+1}dy^2\right)\,.
\end{align}
Finally, we compute the norm on $\Sigma_2$ to be
\begin{equation}
\label{eq:sigma2contrib}
    \left.\langle h^{(m)}, h^{(n)} \rangle \right|_{\Sigma_2}  =|a_m|^2\frac{32\pi^2 r_{\rm n}^2}{\ell_{\rm dS}^2}|m|(m^2-1)\left[1-(-1)^{|m|}(1-2m^2)\right]\delta_{m,n}\,.
\end{equation}
The sum of \eqref{eq:sigma1contrib} and \eqref{eq:sigma2contrib} gives the norm of two arbitrary tensor eigenmodes $h^{(m)}$ and $h^{(n)}$ to be
\begin{equation}
    \left.\langle h^{(m)}, h^{(n)} \rangle \right|_{\mathcal{C}}= |a_m|^2\frac{32\pi^2 r_{\rm n}^2}{\ell_{\rm dS}^2}|m|(m^2-1) ~\delta_{m,n}\,.
\end{equation}
Demanding orthogonality among the zero modes finally gives the value of the normalization to be
\begin{equation}
    |a_m|=\frac{\ell_{\rm dS}}{4\pi r_{\rm n}\sqrt{2|m|(m^2-1)}}\,.
\end{equation}
Further, one can easily check that the zero modes in the analytically continued sector can also be obtained from a set of non-normalizable vector fields following $h_{\mu\nu} = \mathcal{L}_{\xi}\bar{g}_{\mu\nu}$ which are explicitly given by
\begin{equation}
    \xi\left(\tau, y\right) = e^{in\tau}\left(f_{1}\left(y\right)\partial_{y} + f_{2}\left(y\right)\partial_{\tau}\right)\,,
\end{equation}
where
\begin{equation}
    \begin{aligned}
        f_1(y)&= \sqrt{\frac{|n|}{\left(n^2-1\right)}} \left(\frac{y-i}{y+i}\right)^{\frac{\left| n\right| }{2}} (y+i \left| n\right| )\,,\\
        f_2(y)&= \sqrt{\frac{1}{|n|\left(n^2-1\right)}}  \left(\frac{y-i}{y+i}\right)^{\frac{\left| n\right| }{2}}\frac{i  \left(i y \left| n\right| -n^2+y^2+1\right)}{y^2+1}\,. 
    \end{aligned}
\end{equation}
Similar to the boundary diffeomorphism generating vector fields of near-cold black hole (given by \eqref{eq:large_gauge_VF_C_1}), the $n=0, \pm 1$ modes here are also divergent. Overall, the central message here is that even though the emergent dS$_2$ geometry in the near-horizon, near-extremal limit of the Nariai black hole initially was defined for a compact interval of the radial coordinate $y \in (-1,+1)$, one can, and in fact, in the present context of Nariai black hole geometries, \textit{should}  analytically continue to obtain tensor zero modes which happen to be generated by non-normalizable diffeomorphism generators {{at late time $y=iy_b$}}. A similar analysis for the tensor zero modes may also be performed using the Maldacena contour leading to the same results for the norm of the tensor modes as above. The Hartle-Hawking or Maldacena contour in the lower half plane will give us these nontrivial zero modes in the far past. {{We will compute the $\log T$ correction to the wavefunction defined at $y=iy_b$ from these boundary zero modes in the subsequent section.}}

\subsection{Ultracold black hole}\vspace{1em}
We employ the same strategy as before to obtain the graviton tensor zero modes of the ultracold black hole. The extremal spacetime \eqref{eq:NHE_metric_UC} has the topology of a two-dimensional Euclidean plane times a sphere. Let us assume that the zero modes are generated by a set of vector fields
\begin{equation}
\xi\left(\tau, y\right) = e^{i n \tau} \left(f_{1}(y) \partial_{y} + f_{2}(y) \partial_{\tau} \right),
\end{equation}
such that $h_{\mu\nu} = \mathcal{L}_{\xi} \bar{g}_{\mu\nu}$ 
belongs to the kernel of the linearized operator \eqref{eq:Lichnerowicz_op}. The vector fields give rise to the following solution, which can be derived straightforwardly
\begin{equation}
    f_{1}(y) = \frac{c_{1}\, e^{n y} + c_{2}\,e^{-n y}}{n}\,,\quad f_{2}(y) = \frac{i f_{1}'(y)}{n}\,, \label{eq:vector_UC}
\end{equation}
where $c_{1}$ and $c_{2}$ are two arbitrary constants, which should be fixed by normalization. The tensor zero modes themselves can be written as
\begin{equation}\label{eq:zeromodes_UC}
    h_{\mu\nu}dx^{\mu}dx^{\nu} = 2 e^{in\tau} \left(c_{1}\, e^{n y} + c_{2}\, e^{-n y} \right) \left(-d\tau^2 + 2\, i\, \frac{c_{1}\, e^{n y} - c_{2}\, e^{-n y}}{c_{1}\, e^{n y} + c_{2}\, e^{-n y}}\, d\tau dy + dy^2\right)\,.
\end{equation}
This is not the end of the story, though. Recall that the relevant zero modes have to be \emph{normalizable}, whereas the vector fields that generate them must be \emph{non-normalizable}. A simple calculation shows that $h_{\mu\nu}$ has a finite norm when $c_{1} = 0$. However, in this case, the vector fields given by equation \eqref{eq:vector_UC} are also \emph{normalizable}. Thus, they can be continuously connected to the identity operator, rendering the associated gauge transformations trivial. Therefore, the above zero modes do not contribute to the one-loop correction at all.

The question remains if there is no non-trivial zero modes for the Euclidean case. We do not address this question in the present article and leave it as a future work. Our analysis merely establishes that no non-trivial gauge transformation of the given form exists. 
Similar to the Nariai limit, we can consider analytic continuation in time following an appropriate coordinate choice. Presence of a boundary will give nontrivial large gauge modes\footnote{See \cite{GonzalezLezcano:2023cuh, GonzalezLezcano:2023uar, GonzalezLezcano:2024rsi, Correa:2019rdk} for a more general analysis of  similar such occurrences in 2-dimensional non-compact spaces which typically arise at the near horizon geometries of black hole spacetimes.  }. However to move away from extremality in the ultracold limit, one must select an ensemble that permits variations in charge and mass, while ensuring that we remain within the \emph{shark fin}. Hence, the path integral computation requires more attention and investigation.
\section{Logarithmic correction to the extremal entropy} \label{sec:sec4}
Due to the small temperature correction to the spacetime metric  and the field strength, the Lichnerowicz operator also gets corrected. Let us denote the leading order correction to the operator by $\delta\Delta^{\alpha\beta\mu\nu}$. We may take recourse to first order perturbation theory to determine the new eigenvalues and eigenfunctions of this operator. The eigenvalue equation now reads 
\begin{equation}
\left(\Delta^{\alpha\beta\mu\nu}+\delta\Delta^{\alpha\beta\mu\nu}\right)\left(h_{\mu\nu} + \delta h_{\mu\nu} \right) = \left(\lambda_{n} + \delta\lambda_{n} \right) \left(h^{\alpha \beta} + \delta h^{\alpha \beta} \right)\,,
\end{equation}
where $\delta$ terms stand for the first order correction over the orthonormal eigenspectrum $\{\lambda_{n}, h_{\mu\nu}\}$. The first order correction to the eigenvalue can be expressed as \cite{Kapec:2023ruw}
\begin{equation} \label{eq:logT_integral}
    \delta\lambda_{n}\left(T\right) = \int d^4x\, \sqrt{\bar{g}}\, h_{\alpha\beta}^*\, \delta\Delta\left(T\right)^{\alpha\beta\mu\nu}\, h_{\mu\nu}\,.
\end{equation}
Similarly, the partition function, being a Gaussian integral over bosonic fields, changes to
\begin{equation}
    \mathcal{Z} = \prod_n\,\frac{1}{\lambda_{n}+\delta\lambda_{n}\left(T\right)} \implies \log\mathcal{Z} = -\sum_{n} \log\left(\lambda_{n}+\delta\lambda_{n}\left(T\right)\right)\,.
\end{equation}
At this point, let us pause and see how from the above function we could generate terms proportional to $\log T$ which would account for the $\log T$ corrections to the black hole entropy. Clearly for any $\lambda_0 \neq 0$, we could scale the argument of the $\log$ function. This generates a series of the kind $\log(1+f_n(T))$ with $f_n(T) = \tfrac{\delta \lambda_n}{\lambda_n}\ll1$ in appropriate units, recalling that we are working in the perturbative regime where $T\ll 1$. Then we see that these modes generate corrections which are polynomials in the perturbing parameter $T$, namely  
\begin{eqnarray}
    \log(1+f_n(T)) = \sum_{m} c_{n,m} T^m\,.
\end{eqnarray} The crucial point here is that corrections as they no doubt are, they are polynomial in $T$ and subleading to logarithmic corrections. The $\log T$ corrections, therefore, necessarily come from the zero modes $(\lambda_n =0)$ of the Lichnerowicz operator. In this case, the partition function above becomes 
\begin{eqnarray}
     \log\mathcal{Z} = -\sum_{n} \log\left(\delta\lambda_{n}\left(T\right)\right)\,,
\end{eqnarray} which gives the contribution to the $\log$ corrections of the black hole. The upshot of this argument is that \textit{only} the zero modes of the Lichnerowicz operator contribute to the $\log$ corrections, which we aim to evaluate. We now turn to the evaluation of the integral in \eqref{eq:logT_integral} for the cases of cold and Nariai black holes. \vspace{-0.5em}

\subsection{Cold black hole}
For the first case, we find that the relevant integral in equation \eqref{eq:logT_integral} is given by
\begin{equation}
\begin{split}
    \delta\lambda_{n} =- \frac{16 |n| \pi\left(n^2-1\right) T }{  r_0 \left(\ell_4^2-6 r_{0}^2\right)}&\int_1^{\infty} dy\,(y-1)^{|n|-2} (y+1)^{-|n|-4}\\& \Big(\ell_4^2 \left(|n| \left(-|n| (y+2)+y (y+2)-4\right)+4 y-2\right)\\&+r_{0}^2 \left(4 n^2 (y+2)-5 |n| (y-1) (y+3)+2 (y-1)^2 (y+4)\right)\Big)\,.
\end{split}
\end{equation}

The $y$ integration leads to
\begin{equation}\label{eq:logT_eigenvalue_correction_cold}
    \delta\lambda_{n} = \frac{2\pi\,n\,T}{r_{0}}\,,\qquad n \geq 2\,,
\end{equation}
and  the $\log$ of the partition function is
\begin{equation}\label{eq:logT_logZ_cold}
    \delta\log\mathcal{Z} = -\log\left(\prod_{n=2}^{\infty} \frac{2\pi n\,T}{r_{0}} \right)\,.
\end{equation}
There is still an infinite product which could be evaluated using standard regularization technique. We describe this last detail later in this section.

\subsection{Nariai black hole}

The integral expression for the leading order correction to the eigenvalue for the near-extremal Nariai black hole is
\begin{equation}
\begin{split}
    \delta\lambda_{n} = &\frac{16i\pi|n| \left(n^2-1\right) T }{\rn \left(\ell_4^2-6 \rn^2\right)}\int_{0}^{\infty} dy\,(-iy-1)^{|n|-2} (-iy+1)^{-|n|-4}\\& \Big(\ell_4^2 \left(|n| \left(-|n| (-iy+2)-iy (-iy+2)-4\right)-4i y-2\right)\\&+\rn^2 \left(4 n^2 (-iy+2)-5 |n| (-iy-1) (-iy+3)+2 (-iy-1)^2 (-iy+4)\right)\Big)\\& +\frac{16i\pi |n| \left(n^2-1\right) T }{  \rn \left(\ell_4^2-6 \rn^2\right)}\int_1^{0} du\,(u-1)^{|n|-2} (u+1)^{-|n|-4}\\& \Big(\ell_4^2 \left(|n| \left(-|n| (u+2)+u (u+2)-4\right)+4 u-2\right)\\&+\rn^2 \left(4 n^2 (u+2)-5 |n| (u-1) (u+3)+2 (u-1)^2 (u+4)\right)\Big)\,.
\end{split}
\end{equation} 
From here, we find the similar correction to the eigenvalue as in the previous (cold) case
\begin{equation}\label{eq:logT_eigenvalue_correction_nariai}
    \delta\lambda_{n} = \frac{2\pi\,n\,T}{\rn} \,,\quad n \geq 2 \,.
\end{equation}
Hence the correction to the partition function is
\begin{equation}\label{eq:logT_logZ_nariai}
    \delta\log\mathcal{Z} = -\log\left(\prod_{n=2}^{\infty} \frac{2\pi n\,T}{\rn} \right)\,.
\end{equation}
Note that due to involvement of analytic continuation, we are actually computing the quantum correction of the wavefunction at late times.
\subsection{Regularization}
We have obtained the correction to the Euclidean partition function due to linear fluctuation over the near-extremal solution. The results \eqref{eq:logT_logZ_cold} and \eqref{eq:logT_logZ_nariai} involve an infinite product over the integer $n$. To evaluate this product, we use the generalized zeta function method of \cite{Hawking:1976ja}. In this technique, one first constructs a generalized zeta function from the eigenvalues
\begin{equation}\label{gen_zeta}
    \mathfrak{Z}\left(s\right) \equiv \sum_{k=1}^{\infty} \Lambda_{k}^{-s},
\end{equation}
the gradient of the generalized zeta function $\mathfrak{Z}\left(s\right)$ at $s=0$ is formally equal to $-\sum_{k} \log\Lambda_{k}\,$. This gives us
\begin{equation}
    \delta\log \mathcal{Z} = \left.\frac{d\mathfrak{Z}\left(s\right)}{d s}\right|_{s=0}.
\end{equation}
Upon using the eigenvalue expression in equations \eqref{eq:logT_eigenvalue_correction_cold} and \eqref{eq:logT_eigenvalue_correction_nariai}, we finally obtain
\begin{equation}\label{eq:logT_answer}
    \delta\log \mathcal{Z} \sim \frac{3}{2}\log T + T-\mathrm{independent ~constant} \,.
\end{equation}
The constant piece in $\delta \log \mathcal{Z}$ is determined by $r_{0}$ or $\rn$, depending on which near-horizon limit we are concerned about.

\section{Conclusions and future directions} \label{sec:conclude}

In this paper, we have examined the $\log{T}$ corrections to the thermodynamic entropy of a near-extremal asymptotically dS Reissner-Nordstr\"om black hole in four-dimensional spacetime. Unlike charged black holes in flat or AdS spacetimes, the dS Reissner-Nordstr\"om black hole admits three distinct extremal limits due to the presence of three horizons with interesting topology emerging in each of the near-extremal near-horizon limits. While the case where the inner and outer horizons coincide has been well studied, owing to its similarity to the flat and AdS black hole cases, we focus on the more complex Nariai limit, where the outer and cosmological horizons merge. In the near-horizon limit, due to the S$^2$ topology of the Euclidean dS$_2$ space, one might naively expect the absence of nontrivial graviton tensor zero modes supported by the kinetic operator in the path integral away from extremality, which includes the perturbative $T$ corrections. However, implementing an analytic continuation in the radial coordinate at the equator to transition into a dS$_2$ spacetime with (1,1) signature allows us to compute the graviton (tensor) zero modes. The analytic continuation allows us to make the radial coordinate non-compact thus allowing non-normalizable diffeomorphism generators giving rise to normalizable tensor zero modes. These are precisely the tensor zero modes that contributed to $\frac{3}{2}\log T$ corrections to the low-temperature thermodynamics.

Perhaps the most intricate and non-trivial among all the three extremal limits is the case of ultracold black hole. Firstly, unlike the other two cases, this extremal limit is a point in the $(M,Q)$ parameter space thus making the black hole move away from extremality significantly more difficult. As pointed in \cite{Castro:2022cuo}, deviations from extremality does not exactly \emph{heat up} the $\mathbb{E}^2$ part. Infact, \cite{Castro:2022cuo} goes into detailing exactly how one should give a \emph{kick} to the ultracold black hole in order to work in an ensemble where both the charge and mass vary keeping one inside the \emph{shark fin}. It will be interesting to explore the $\log T$ corrections to entropy staying within such an ensemble but we relegate that analysis for future work. Secondly, the analysis for determining the large gauge modes which gives rise to normalizable tensor zero modes will require one to impose nontrivial fall-off conditions on the metric, in a \emph{Bondi like}-patch after defining null coordinate $u \equiv \tau-y$ and expressing the $\mathbb{E}^2 \times S^2$ metric properly. It is precisely these intricacies which makes the $\log T$ correction in the context of ultracold black holes significantly more complicated. We intend to explore this further in future work.

Another crucial limitation of the current work is the fact that we are only focussed on corrections arising from tensor zero modes. Of course there will be corrections arising from the scalar as well as vector sectors. Whether these will have \emph{universal} behaviour akin to the tensor modes is indeed an interesting issue to address. Finally, an obvious question following our current investigation is to look for $\log T$ corrections to entropy for other black holes in dS spacetime (involving rotating cases and more general configurations) and also investigate such corrections for higher dimensional black holes.

\section*{Acknowledgement}
The authors thank Sunil Kumar Sake for multiple discussions involving gravitational path integrals in dS$_2$. The authors also thank Imtak Jeon, Seok Kim, Alfredo G. Lezcano and Leopoldo A. Pando Zayas for various discussions related to black hole physics. The authors thank Adam R. Levine, Matthew M. Roberts and Akhil Sivakumar for their comments on an earlier draft of the manuscript. The work of A.M. is supported by POSTECH BK21 postdoctoral fellowship. The work of D.M. is supported by the \emph{Young Scientist Training (YST) Fellowship} from Asia Pacific Center for Theoretical Physics. A.M. and D.M. acknowledge support by the National Research Foundation of Korea (NRF) grant funded by the Korean government (MSIT) (No. 2022R1A2C1003182). A.R. is supported by the National Research Foundation of Korea
(NRF) Grant 2021R1A2C2012350. A.M and D.M also thank Yukawa Institute for Theoretical Physics at Kyoto University, where a part of this work was done during the YITP workshop: Quantum Gravity and Information in Expanding Universe (YITP-W-24-19).

\appendix
\section{Near-extremal expansion for cold vs ultracold black holes}\label{appendix:A}

In this section, we elucidate and motivate the power series expansion in the near ultracold limit and contrast it with the near cold and near Nariai limits. For the near cold black hole (outer and inner horizons i.e. $r_+$ and $r_{-}$ are very close to each other), we start by assuming a power series expansion of $r_{\pm}$ in an arbitrary infinitesimal parameter $\lambda$
\begin{align}
    r_{+} = r_{0} + \alpha\,\lambda\ \quad\,,\quad r_{-} = r_{0} + \beta\,\lambda\,,
\end{align}
which results in the following expansion of $r_{c}$ using equation \eqref{eq:curvature_and_rc}
\begin{equation}
    r_{c} = -r_{0} + \sqrt{\ell_4^2 - 2r_{0}^2} - \frac{1}{2}\left(\alpha+\beta\right)\left(1+\frac{2r_{0}}{\sqrt{\ell_4^2 - 2r_{0}^2}}\right)\lambda\,.
\end{equation}
Eventually, \eqref{eq:temperature_outer} gives the series expansion of $T_{+}$ as
\begin{equation}
    T_{+} = \frac{\left(\ell_4^2 - 6 r_{0}^2\right)\left(\alpha-\beta\right)}{4\pi \ell_4^2 r_{0}^2}\lambda\,.
\end{equation}
Since we see in the above that $T_{+} \sim \lambda$ i.e. they are of similar order, one may as well use $T_{+}$ itself as the expansion parameter. The same conclusion holds for the Nariai black hole.

Near the ultracold limit we assume
\begin{align}
    r_{+} = \ruc + \alpha\,\lambda\ \quad\,,\quad\ r_{-} = \ruc + \beta\,\lambda\,,
\end{align}
which implies
\begin{equation}
\begin{split}
    r_{c} &= -\ruc + \sqrt{\ell_{4}^2 - 2\ruc^2} - \frac{\left(2\ruc + \sqrt{\ell_{4}^2 - 2\ruc^2}\right) \left(\alpha+\beta\right)}{2\sqrt{\ell_{4}^2 - 2\ruc^2}} \lambda\,,\\ \text{and}\quad
    T_{+} &= -\frac{\left(\ell_{4}^2 - 6\ruc^2 \right)\left(\alpha-\beta\right)}{4\pi\ell_4^2\ruc^2}\lambda^2\, .
\end{split}
\end{equation}
In this case, we see $\lambda \sim \sqrt{T_{+}}$ which justifies our expansion in $\sqrt{T}$ in section \ref{sec:sec3_subsec3}.

\section{On-shell action and wavefunction in Nariai limit}
\label{appendix:B}
In this section, we evaluate the leading partition function for Nariai black holes prior to near-extremal expansion. Thus, we are sitting at exact extremality and applying the saddle-point approximation. We will be evaluating the leading behaviour of the partition function for near-horizon Nariai black holes described by the metric \eqref{eq:NHE_metric_N} where the path integral is evaluated along the contour $\Sigma_1 \cup \Sigma_2$ (colored brown) in Fig. \ref{fig:eefig}. The contour is first real running from $y=1$ to $y=0$. Subsequently, it becomes purely imaginary and runs along $y=i0^{+}$ to $y=iy_b$ where $y_b$ is a large real number representing the boundary regulator.\\
For the part of the contour $\Sigma_1$ (refer to Fig. \ref{fig:eefig}), we have $F_{\tau y}=\frac{ir_{\rm n}\left(1-\frac{3r_{\rm n}^2}{\ell_4^2} \right)^{1/2}}{\frac{6r_{\rm n}^2}{\ell_4^2}-1}$, while for $\Sigma_2$, we have $F_{\tau y}=-\frac{r_{\rm n}\left(1-\frac{3r_{\rm n}^2}{\ell_4^2} \right)^{1/2}}{\frac{6r_{\rm n}^2}{\ell_4^2}-1}$. The Ricci scalar as well as the square of the gauge field strength $F^2$ for either branch $\Sigma_1$ and $\Sigma_2$ of the contour $\mathcal{C}$ is the same and can be evaluated to be
\begin{equation}
     R= \frac{12}{\ell_4^2} \quad ,\quad F^2= \frac{6}{\ell_4^2}-\frac{2}{r_{\rm n}^2}\,.
\end{equation}
The 4-dimensional bulk action evaluated along the contour $\mathcal{C} \equiv \Sigma_1 \cup \Sigma_2$ finally gives
\begin{equation}
    I^{(4D)}_{\rm bulk}= \frac{\pi r_{\rm n}^2}{G_N\left(\frac{6r_{\rm n}^2}{\ell_4^2}-1 \right)}(-1+ iy_b)=\frac{\pi \ell_{\rm{dS}}^2}{G_N}(iy_b-1)\,,
\end{equation}
where $\ell_{\text{dS}}$ is the radius of the two dimensional de Sitter geometry, given in equation \eqref{eq:ds_curvature}. The boundary action is given by 
\begin{equation}
    \mathcal{S}_{\rm bdy}^{(3D)}=\frac{1}{16 \pi G_N}\int_{\partial \mathcal{M}}d^3x\ 2\sqrt{\gamma}K + \frac{1}{4 \pi G_N}\int_{\partial \mathcal{M}}d^3x\ \sqrt{\gamma}F^{\mu \nu}n_{\mu}A_{\nu}\,.
\end{equation}
Note here the boundary contribution to the Maxwell action is derived by adding a boundary term which puts the electric solution in the fixed charge (canonical) ensemble \cite{GotoThesis}. Note that the above action is evaluated on the \emph{boundary} i.e. the point $y=iy_b$ in the complex $y$-plane. Infact, from Fig. \ref{fig:eefig}, it follows that the boundary metric is explicitly given by
\begin{equation}
    ds^2_{\rm bdy}= \ell_{\rm dS}^2\left(1+y_b^2\right) d\tau^2 + r_{\mathrm{n}}^2\, d\Omega_2^2\,.
\end{equation}
with $\tau$ circle having radius $\ell_{\rm{dS}}\sqrt{1+y_b^2}$. The outward pointing normal at $y=iy_b$ is given by
\begin{equation}
    n^y = \frac{1}{\ell_{\rm dS}}\sqrt{1+y_b^2}\,.
\end{equation}
This eventually leads to
\begin{equation}
   2\sqrt{\gamma}K = -2r_{\rm n}^2y_b \sin \theta\ \quad\ \text{and}\quad \sqrt{\gamma}F^{\mu \nu}n_{\mu}A_{\nu} = \ell_{\rm dS}^2y_b \left( 1-\frac{3r_{\rm n}^2}{\ell_4^2}\right)\sin \theta\,.
\end{equation}
Finally, the boundary piece (after the $\tau, \theta$ and $\phi$ integral) leads to
\begin{equation}
     \left. {\mathcal{S}}_{\rm bdy}^{(3D)}\right|_{\rm on-shell} =\frac{\pi \ell_{\rm{dS}}^2}{G_N}y_b \left(1-\frac{2\rn^2}{\ell_{\rm{dS}}^2}\right)\,,
\end{equation}
which eventually gives the total on-shell action
\begin{equation}
   {\cal{S}}_{\rm{on-shell}}= \frac{\pi \ell_{\rm dS}^2}{G_N}(iy_b-1)+\frac{\pi \ell_{\rm dS}^2}{G_N}y_b \left(1-\frac{2r_{\rm n}^2}{\ell_{\rm dS}^2} \right)\,.
\end{equation}
The leading Hartle-Hawking wave function can be obtained from $e^{i{\cal{S}}_{\rm on-shell}}$. Note that in the conventional path integral framework, ${\cal{S}}_{\rm on-shell}$ is equivalent to $S_0/2$, while $S_0$ being the thermodynamic entropy. In the above, the leading behaviour of the on-shell action as $Q \rightarrow 0$ i.e. $3r_{\rm n}^2 \rightarrow \ell_4^2$ sans the cut-off dependent part boils down to the leading part of $-S_c$ as given by \eqref{eq:NariaiCosmoEntropy}.


\bibliography{ref.bib}

\providecommand{\href}[2]{#2}\begingroup\raggedright\begin{thebibliography}{10}

\bibitem{PhysRevD.7.2333}
J.~D. Bekenstein, \emph{Black holes and entropy}, \href{https://doi.org/10.1103/PhysRevD.7.2333}{\emph{Phys. Rev. D} {\bfseries 7} (1973) 2333}.

\bibitem{Hawking:1975vcx}
S.~W. Hawking, \emph{{Particle Creation by Black Holes}}, \href{https://doi.org/10.1007/BF02345020}{\emph{Commun. Math. Phys.} {\bfseries 43} (1975) 199}.

\bibitem{Solodukhin:1994yz}
S.~N. Solodukhin, \emph{{The Conical singularity and quantum corrections to entropy of black hole}}, \href{https://doi.org/10.1103/PhysRevD.51.609}{\emph{Phys. Rev. D} {\bfseries 51} (1995) 609} [\href{https://arxiv.org/abs/hep-th/9407001}{{\ttfamily hep-th/9407001}}].

\bibitem{Fursaev:1994te}
D.~V. Fursaev, \emph{{Temperature and entropy of a quantum black hole and conformal anomaly}}, \href{https://doi.org/10.1103/PhysRevD.51.R5352}{\emph{Phys. Rev. D} {\bfseries 51} (1995) 5352} [\href{https://arxiv.org/abs/hep-th/9412161}{{\ttfamily hep-th/9412161}}].

\bibitem{Mann:1997hm}
R.~B. Mann and S.~N. Solodukhin, \emph{{Universality of quantum entropy for extreme black holes}}, \href{https://doi.org/10.1016/S0550-3213(98)00094-7}{\emph{Nucl. Phys. B} {\bfseries 523} (1998) 293} [\href{https://arxiv.org/abs/hep-th/9709064}{{\ttfamily hep-th/9709064}}].

\bibitem{Kaul:2000kf}
R.~K. Kaul and P.~Majumdar, \emph{{Logarithmic correction to the Bekenstein-Hawking entropy}}, \href{https://doi.org/10.1103/PhysRevLett.84.5255}{\emph{Phys. Rev. Lett.} {\bfseries 84} (2000) 5255} [\href{https://arxiv.org/abs/gr-qc/0002040}{{\ttfamily gr-qc/0002040}}].

\bibitem{Carlip:2000nv}
S.~Carlip, \emph{{Logarithmic corrections to black hole entropy from the Cardy formula}}, \href{https://doi.org/10.1088/0264-9381/17/20/302}{\emph{Class. Quant. Grav.} {\bfseries 17} (2000) 4175} [\href{https://arxiv.org/abs/gr-qc/0005017}{{\ttfamily gr-qc/0005017}}].

\bibitem{Solodukhin:2010pk}
S.~N. Solodukhin, \emph{{Entanglement entropy of round spheres}}, \href{https://doi.org/10.1016/j.physletb.2010.09.018}{\emph{Phys. Lett. B} {\bfseries 693} (2010) 605} [\href{https://arxiv.org/abs/1008.4314}{{\ttfamily 1008.4314}}].

\bibitem{Banerjee:2010qc}
S.~Banerjee, R.~K. Gupta and A.~Sen, \emph{{Logarithmic Corrections to Extremal Black Hole Entropy from Quantum Entropy Function}}, \href{https://doi.org/10.1007/JHEP03(2011)147}{\emph{JHEP} {\bfseries 03} (2011) 147} [\href{https://arxiv.org/abs/1005.3044}{{\ttfamily 1005.3044}}].

\bibitem{Banerjee:2011jp}
S.~Banerjee, R.~K. Gupta, I.~Mandal and A.~Sen, \emph{{Logarithmic Corrections to N=4 and N=8 Black Hole Entropy: A One Loop Test of Quantum Gravity}}, \href{https://doi.org/10.1007/JHEP11(2011)143}{\emph{JHEP} {\bfseries 11} (2011) 143} [\href{https://arxiv.org/abs/1106.0080}{{\ttfamily 1106.0080}}].

\bibitem{Sen:2012kpz}
A.~Sen, \emph{{Logarithmic Corrections to N=2 Black Hole Entropy: An Infrared Window into the Microstates}}, \href{https://doi.org/10.1007/s10714-012-1336-5}{\emph{Gen. Rel. Grav.} {\bfseries 44} (2012) 1207} [\href{https://arxiv.org/abs/1108.3842}{{\ttfamily 1108.3842}}].

\bibitem{Sen:2012cj}
A.~Sen, \emph{{Logarithmic Corrections to Rotating Extremal Black Hole Entropy in Four and Five Dimensions}}, \href{https://doi.org/10.1007/s10714-012-1373-0}{\emph{Gen. Rel. Grav.} {\bfseries 44} (2012) 1947} [\href{https://arxiv.org/abs/1109.3706}{{\ttfamily 1109.3706}}].

\bibitem{Strominger:1996sh}
A.~Strominger and C.~Vafa, \emph{{Microscopic origin of the Bekenstein-Hawking entropy}}, \href{https://doi.org/10.1016/0370-2693(96)00345-0}{\emph{Phys. Lett. B} {\bfseries 379} (1996) 99} [\href{https://arxiv.org/abs/hep-th/9601029}{{\ttfamily hep-th/9601029}}].

\bibitem{2013JHEP...04..156S}
A.~{Sen}, \emph{{Logarithmic corrections to Schwarzschild and other non-extremal black hole entropy in different dimensions}}, \href{https://doi.org/10.1007/JHEP04(2013)156}{\emph{JHEP} {\bfseries 2013} (2013) 156} [\href{https://arxiv.org/abs/1205.0971}{{\ttfamily 1205.0971}}].

\bibitem{Bhattacharyya:2012bps}
S.~{Bhattacharyya}, B.~{Panda} and A.~{Sen}, \emph{{Heat kernel expansion and extremal Kerr-Newmann black hole entropy in Einstein-Maxwell theory}}, \href{https://doi.org/10.1007/JHEP08(2012)084}{\emph{JHEP} {\bfseries 2012} (2012) 84} [\href{https://arxiv.org/abs/1204.4061}{{\ttfamily 1204.4061}}].

\bibitem{Banerjee:2020bkp}
G.~{Banerjee}, S.~{Karan} and B.~{Panda}, \emph{{Logarithmic correction to the entropy of extremal black holes in N = 1 Einstein-Maxwell supergravity}}, \href{https://doi.org/10.1007/JHEP01(2021)090}{\emph{JHEP} {\bfseries 2021} (2021) 90} [\href{https://arxiv.org/abs/2007.11497}{{\ttfamily 2007.11497}}].

\bibitem{Karan:2020kpx}
S.~{Karan} and B.~{Panda}, \emph{{Logarithmic corrections to black hole entropy in matter coupled N {\ensuremath{\geq}} 1 Einstein-Maxwell supergravity}}, \href{https://doi.org/10.1007/JHEP05(2021)104}{\emph{JHEP} {\bfseries 2021} (2021) 104} [\href{https://arxiv.org/abs/2012.12227}{{\ttfamily 2012.12227}}].

\bibitem{Karan:2021kpx}
S.~{Karan} and B.~{Panda}, \emph{{Generalized Einstein-Maxwell theory: Seeley-DeWitt coefficients and logarithmic corrections to the entropy of extremal and nonextremal black holes}}, \href{https://doi.org/10.1103/PhysRevD.104.046010}{\emph{Phys. Rev. D} {\bfseries 104} (2021) 046010} [\href{https://arxiv.org/abs/2104.06381}{{\ttfamily 2104.06381}}].

\bibitem{Banerjee:2021bpx}
G.~{Banerjee} and B.~{Panda}, \emph{{Logarithmic corrections to the entropy of non-extremal black holes in N = 1 Einstein-Maxwell supergravity}}, \href{https://doi.org/10.1007/JHEP11(2021)214}{\emph{JHEP} {\bfseries 2021} (2021) 214} [\href{https://arxiv.org/abs/2109.04407}{{\ttfamily 2109.04407}}].

\bibitem{Bhattacharyya:2012bgm}
S.~{Bhattacharyya}, A.~{Grassi}, M.~{Marino} and A.~{Sen}, \emph{{A One-Loop Test of Quantum Supergravity}}, \href{https://doi.org/10.48550/arXiv.1210.6057}{\emph{arXiv e-prints} (2012) arXiv:1210.6057} [\href{https://arxiv.org/abs/1210.6057}{{\ttfamily 1210.6057}}].

\bibitem{Liu:2017prl}
J.~T. Liu, L.~A. Pando~Zayas, V.~Rathee and W.~Zhao, \emph{{One-Loop Test of Quantum Black Holes in anti{\textendash}de Sitter Space}}, \href{https://doi.org/10.1103/PhysRevLett.120.221602}{\emph{Phys. Rev. Lett.} {\bfseries 120} (2018) 221602} [\href{https://arxiv.org/abs/1711.01076}{{\ttfamily 1711.01076}}].

\bibitem{Liu:2017lpr}
J.~T. {Liu}, L.~A. {Pando Zayas}, V.~{Rathee} and W.~{Zhao}, \emph{{Toward microstate counting beyond large N in localization and the dual one-loop quantum supergravity}}, \href{https://doi.org/10.1007/JHEP01(2018)026}{\emph{JHEP} {\bfseries 2018} (2018) 26} [\href{https://arxiv.org/abs/1707.04197}{{\ttfamily 1707.04197}}].

\bibitem{Gang:2019gnp}
D.~{Gang}, N.~{Kim} and L.~A. {Pando Zayas}, \emph{{Precision microstate counting for the entropy of wrapped M5-branes}}, \href{https://doi.org/10.1007/JHEP03(2020)164}{\emph{JHEP} {\bfseries 2020} (2020) 164} [\href{https://arxiv.org/abs/1905.01559}{{\ttfamily 1905.01559}}].

\bibitem{Pando:2019tti}
L.~A. {Pando Zayas} and Y.~{Xin}, \emph{{Topologically twisted index in the 't Hooft limit and the dual AdS$_{4}$ black hole entropy}}, \href{https://doi.org/10.1103/PhysRevD.100.126019}{\emph{Phys. Rev. D} {\bfseries 100} (2019) 126019} [\href{https://arxiv.org/abs/1908.01194}{{\ttfamily 1908.01194}}].

\bibitem{Benini:2020bgp}
F.~{Benini}, D.~{Gang} and L.~A. {Pando Zayas}, \emph{{Rotating black hole entropy from M5-branes}}, \href{https://doi.org/10.1007/JHEP03(2020)057}{\emph{JHEP} {\bfseries 2020} (2020) 57} [\href{https://arxiv.org/abs/1909.11612}{{\ttfamily 1909.11612}}].

\bibitem{Pando:2020pzx}
L.~A. {Pando Zayas} and Y.~{Xin}, \emph{{Universal logarithmic behavior in microstate counting and the dual one-loop entropy of AdS$_{4}$ black holes}}, \href{https://doi.org/10.1103/PhysRevD.103.026003}{\emph{Phys. Rev. D} {\bfseries 103} (2021) 026003} [\href{https://arxiv.org/abs/2008.03239}{{\ttfamily 2008.03239}}].

\bibitem{Hristov:2021hrx}
K.~{Hristov} and V.~{Reys}, \emph{{Factorization of log-corrections in AdS$_{4}$/CFT$_{3}$ from supergravity localization}}, \href{https://doi.org/10.1007/JHEP12(2021)031}{\emph{JHEP} {\bfseries 2021} (2021) 31} [\href{https://arxiv.org/abs/2107.12398}{{\ttfamily 2107.12398}}].

\bibitem{David:2021dgn}
M.~{David}, A.~{Gonz{\'a}lez Lezcano}, J.~{Nian} and L.~A. {Pando Zayas}, \emph{{Logarithmic corrections to the entropy of rotating black holes and black strings in AdS$_{5}$}}, \href{https://doi.org/10.1007/JHEP04(2022)160}{\emph{JHEP} {\bfseries 2022} (2022) 160} [\href{https://arxiv.org/abs/2106.09730}{{\ttfamily 2106.09730}}].

\bibitem{David:2021dgl}
M.~{David}, V.~{Godet}, Z.~{Liu} and L.~A. {Pando Zayas}, \emph{{Non-topological logarithmic corrections in minimal gauged supergravity}}, \href{https://doi.org/10.1007/JHEP08(2022)043}{\emph{JHEP} {\bfseries 2022} (2022) 43} [\href{https://arxiv.org/abs/2112.09444}{{\ttfamily 2112.09444}}].

\bibitem{Karan:2022ksx}
S.~{Karan} and G.~S. {Punia}, \emph{{Logarithmic correction to black hole entropy in universal low-energy string theory models}}, \href{https://doi.org/10.1007/JHEP03(2023)028}{\emph{JHEP} {\bfseries 2023} (2023) 28} [\href{https://arxiv.org/abs/2210.16230}{{\ttfamily 2210.16230}}].

\bibitem{Bobev:2023bdh}
N.~{Bobev}, M.~{David}, J.~{Hong}, V.~{Reys} and X.~{Zhang}, \emph{{A compendium of logarithmic corrections in AdS/CFT}}, \href{https://doi.org/10.1007/JHEP04(2024)020}{\emph{JHEP} {\bfseries 2024} (2024) 20} [\href{https://arxiv.org/abs/2312.08909}{{\ttfamily 2312.08909}}].

\bibitem{Karan:2024ksb}
S.~{Karan}, G.~{Singh Punia} and S.~{Biswas}, \emph{{Logarithmic correction to the entropy of black holes in STU supergravity}}, \href{https://doi.org/10.48550/arXiv.2403.11823}{\emph{arXiv e-prints} (2024) arXiv:2403.11823} [\href{https://arxiv.org/abs/2403.11823}{{\ttfamily 2403.11823}}].

\bibitem{Preskill:1991tb}
J.~Preskill, P.~Schwarz, A.~D. Shapere, S.~Trivedi and F.~Wilczek, \emph{{Limitations on the statistical description of black holes}}, \href{https://doi.org/10.1142/S0217732391002773}{\emph{Mod. Phys. Lett. A} {\bfseries 6} (1991) 2353}.

\bibitem{Iliesiu:2020qvm}
L.~V. Iliesiu and G.~J. Turiaci, \emph{{The statistical mechanics of near-extremal black holes}}, \href{https://doi.org/10.1007/JHEP05(2021)145}{\emph{JHEP} {\bfseries 05} (2021) 145} [\href{https://arxiv.org/abs/2003.02860}{{\ttfamily 2003.02860}}].

\bibitem{Iliesiu:2022onk}
L.~V. Iliesiu, S.~Murthy and G.~J. Turiaci, \emph{{Revisiting the Logarithmic Corrections to the Black Hole Entropy}},  \href{https://arxiv.org/abs/2209.13608}{{\ttfamily 2209.13608}}.

\bibitem{Banerjee:2023quv}
N.~Banerjee and M.~Saha, \emph{{Revisiting leading quantum corrections to near extremal black hole thermodynamics}}, \href{https://doi.org/10.1007/JHEP07(2023)010}{\emph{JHEP} {\bfseries 07} (2023) 010} [\href{https://arxiv.org/abs/2303.12415}{{\ttfamily 2303.12415}}].

\bibitem{Kapec:2023ruw}
D.~Kapec, A.~Sheta, A.~Strominger and C.~Toldo, \emph{{Logarithmic Corrections to Kerr Thermodynamics}},  \href{https://arxiv.org/abs/2310.00848}{{\ttfamily 2310.00848}}.

\bibitem{Rakic:2023vhv}
I.~Rakic, M.~Rangamani and G.~J. Turiaci, \emph{{Thermodynamics of the near-extremal Kerr spacetime}},  \href{https://arxiv.org/abs/2310.04532}{{\ttfamily 2310.04532}}.

\bibitem{Banerjee:2023gll}
N.~Banerjee, M.~Saha and S.~Srinivasan, \emph{{Logarithmic corrections for near-extremal black holes}}, \href{https://doi.org/10.1007/JHEP02(2024)077}{\emph{JHEP} {\bfseries 2024} (2024) 077} [\href{https://arxiv.org/abs/2311.09595}{{\ttfamily 2311.09595}}].

\bibitem{Maulik:2024dwq}
S.~Maulik, L.~A. Pando~Zayas, A.~Ray and J.~Zhang, \emph{{Universality in logarithmic temperature corrections to near-extremal rotating black hole thermodynamics in various dimensions}}, \href{https://doi.org/10.1007/JHEP06(2024)034}{\emph{JHEP} {\bfseries 06} (2024) 034} [\href{https://arxiv.org/abs/2401.16507}{{\ttfamily 2401.16507}}].

\bibitem{Kolanowski:2024zrq}
M.~Kolanowski, D.~Marolf, I.~Rakic, M.~Rangamani and G.~J. Turiaci, \emph{{Looking at extremal black holes from very far away}},  \href{https://arxiv.org/abs/2409.16248}{{\ttfamily 2409.16248}}.

\bibitem{Kapec:2024zdj}
D.~Kapec, Y.~T.~A. Law and C.~Toldo, \emph{{Quasinormal Corrections to Near-Extremal Black Hole Thermodynamics}},  \href{https://arxiv.org/abs/2409.14928}{{\ttfamily 2409.14928}}.

\bibitem{Modak:2025gvp}
A.~Modak, A.~Singh and B.~Panda, \emph{{Logarithmic Corrections for Near-extremal Kerr-Newman Black Holes}},  \href{https://arxiv.org/abs/2502.18173}{{\ttfamily 2502.18173}}.

\bibitem{Arnaudo:2024bbd}
P.~Arnaudo, G.~Bonelli and A.~Tanzini, \emph{{One-loop corrections to near extremal Kerr thermodynamics from semiclassical Virasoro blocks}},  \href{https://arxiv.org/abs/2412.16057}{{\ttfamily 2412.16057}}.

\bibitem{Witten:2001kn}
E.~Witten, \emph{{Quantum gravity in de Sitter space}},  in \emph{{Strings 2001: International Conference}}, 6, 2001, \href{https://arxiv.org/abs/hep-th/0106109}{{\ttfamily hep-th/0106109}}.

\bibitem{Strominger:2001pn}
A.~Strominger, \emph{{The dS / CFT correspondence}}, \href{https://doi.org/10.1088/1126-6708/2001/10/034}{\emph{JHEP} {\bfseries 10} (2001) 034} [\href{https://arxiv.org/abs/hep-th/0106113}{{\ttfamily hep-th/0106113}}].

\bibitem{Goheer:2002vf}
N.~Goheer, M.~Kleban and L.~Susskind, \emph{{The Trouble with de Sitter space}}, \href{https://doi.org/10.1088/1126-6708/2003/07/056}{\emph{JHEP} {\bfseries 07} (2003) 056} [\href{https://arxiv.org/abs/hep-th/0212209}{{\ttfamily hep-th/0212209}}].

\bibitem{Kachru:2003aw}
S.~Kachru, R.~Kallosh, A.~D. Linde and S.~P. Trivedi, \emph{{De Sitter vacua in string theory}}, \href{https://doi.org/10.1103/PhysRevD.68.046005}{\emph{Phys. Rev. D} {\bfseries 68} (2003) 046005} [\href{https://arxiv.org/abs/hep-th/0301240}{{\ttfamily hep-th/0301240}}].

\bibitem{Castro:2011xb}
A.~Castro, N.~Lashkari and A.~Maloney, \emph{{A de Sitter Farey Tail}}, \href{https://doi.org/10.1103/PhysRevD.83.124027}{\emph{Phys. Rev. D} {\bfseries 83} (2011) 124027} [\href{https://arxiv.org/abs/1103.4620}{{\ttfamily 1103.4620}}].

\bibitem{Basu:2019mjo}
R.~Basu and A.~Ray, \emph{{Supersymmetric Localization on dS: Sum over topologies}}, \href{https://doi.org/10.1140/epjc/s10052-020-08463-0}{\emph{Eur. Phys. J. C} {\bfseries 80} (2020) 885} [\href{https://arxiv.org/abs/1911.07480}{{\ttfamily 1911.07480}}].

\bibitem{Gibbons:1977mu}
G.~W. Gibbons and S.~W. Hawking, \emph{{Cosmological Event Horizons, Thermodynamics, and Particle Creation}}, \href{https://doi.org/10.1103/PhysRevD.15.2738}{\emph{Phys. Rev. D} {\bfseries 15} (1977) 2738}.

\bibitem{Bekenstein:1972tm}
J.~D. Bekenstein, \emph{{Black holes and the second law}}, \href{https://doi.org/10.1007/BF02757029}{\emph{Lett. Nuovo Cim.} {\bfseries 4} (1972) 737}.

\bibitem{Bousso:2000nf}
R.~Bousso, \emph{{Positive vacuum energy and the N bound}}, \href{https://doi.org/10.1088/1126-6708/2000/11/038}{\emph{JHEP} {\bfseries 11} (2000) 038} [\href{https://arxiv.org/abs/hep-th/0010252}{{\ttfamily hep-th/0010252}}].

\bibitem{Bousso:2000md}
R.~Bousso, \emph{{Bekenstein bounds in de Sitter and flat space}}, \href{https://doi.org/10.1088/1126-6708/2001/04/035}{\emph{JHEP} {\bfseries 04} (2001) 035} [\href{https://arxiv.org/abs/hep-th/0012052}{{\ttfamily hep-th/0012052}}].

\bibitem{Romans:1991nq}
L.~J. Romans, \emph{{Supersymmetric, cold and lukewarm black holes in cosmological Einstein-Maxwell theory}}, \href{https://doi.org/10.1016/0550-3213(92)90684-4}{\emph{Nucl. Phys. B} {\bfseries 383} (1992) 395} [\href{https://arxiv.org/abs/hep-th/9203018}{{\ttfamily hep-th/9203018}}].

\bibitem{Mann:1995vb}
R.~B. Mann and S.~F. Ross, \emph{{Cosmological production of charged black hole pairs}}, \href{https://doi.org/10.1103/PhysRevD.52.2254}{\emph{Phys. Rev. D} {\bfseries 52} (1995) 2254} [\href{https://arxiv.org/abs/gr-qc/9504015}{{\ttfamily gr-qc/9504015}}].

\bibitem{Booth:1998gf}
I.~S. Booth and R.~B. Mann, \emph{{Cosmological pair production of charged and rotating black holes}}, \href{https://doi.org/10.1016/S0550-3213(98)00756-1}{\emph{Nucl. Phys. B} {\bfseries 539} (1999) 267} [\href{https://arxiv.org/abs/gr-qc/9806056}{{\ttfamily gr-qc/9806056}}].

\bibitem{Nariai:1950srt}
H.~{Nariai}, \emph{{On some static solutions of Einstein's gravitational field equations in a spherically symmetric case}}, {\emph{Sci. Rep. Tohoku Univ. Eighth Ser.} {\bfseries 34} (1950) 160}.

\bibitem{Nariai:1999iok}
H.~Nariai, \emph{{On a New Cosmological Solution of Einstein's Field Equations of Gravitation}}, \href{https://doi.org/10.1023/A:1026602724948}{\emph{Gen. Rel. Grav.} {\bfseries 31} (1999) 963}.

\bibitem{Castro:2022cuo}
A.~Castro, F.~Mariani and C.~Toldo, \emph{{Near-extremal limits of de Sitter black holes}}, \href{https://doi.org/10.1007/JHEP07(2023)131}{\emph{JHEP} {\bfseries 07} (2023) 131} [\href{https://arxiv.org/abs/2212.14356}{{\ttfamily 2212.14356}}].

\bibitem{Nayak:2018qej}
P.~Nayak, A.~Shukla, R.~M. Soni, S.~P. Trivedi and V.~Vishal, \emph{{On the Dynamics of Near-Extremal Black Holes}}, \href{https://doi.org/10.1007/JHEP09(2018)048}{\emph{JHEP} {\bfseries 09} (2018) 048} [\href{https://arxiv.org/abs/1802.09547}{{\ttfamily 1802.09547}}].

\bibitem{Almheiri:2014cka}
A.~Almheiri and J.~Polchinski, \emph{{Models of AdS$_{2}$ backreaction and holography}}, \href{https://doi.org/10.1007/JHEP11(2015)014}{\emph{JHEP} {\bfseries 11} (2015) 014} [\href{https://arxiv.org/abs/1402.6334}{{\ttfamily 1402.6334}}].

\bibitem{Maldacena:2019cbz}
J.~Maldacena, G.~J. Turiaci and Z.~Yang, \emph{{Two dimensional Nearly de Sitter gravity}}, \href{https://doi.org/10.1007/JHEP01(2021)139}{\emph{JHEP} {\bfseries 01} (2021) 139} [\href{https://arxiv.org/abs/1904.01911}{{\ttfamily 1904.01911}}].

\bibitem{Moitra:2022glw}
U.~Moitra, S.~K. Sake and S.~P. Trivedi, \emph{{Aspects of Jackiw-Teitelboim gravity in Anti-de Sitter and de Sitter spacetime}}, \href{https://doi.org/10.1007/JHEP06(2022)138}{\emph{JHEP} {\bfseries 06} (2022) 138} [\href{https://arxiv.org/abs/2202.03130}{{\ttfamily 2202.03130}}].

\bibitem{Afshar:2019axx}
H.~Afshar, H.~A. Gonz\'alez, D.~Grumiller and D.~Vassilevich, \emph{{Flat space holography and the complex Sachdev-Ye-Kitaev model}}, \href{https://doi.org/10.1103/PhysRevD.101.086024}{\emph{Phys. Rev. D} {\bfseries 101} (2020) 086024} [\href{https://arxiv.org/abs/1911.05739}{{\ttfamily 1911.05739}}].

\bibitem{Hartle:1983ai}
J.~B. Hartle and S.~W. Hawking, \emph{{Wave Function of the Universe}}, \href{https://doi.org/10.1103/PhysRevD.28.2960}{\emph{Phys. Rev. D} {\bfseries 28} (1983) 2960}.

\bibitem{Moitra:2021uiv}
U.~Moitra, S.~K. Sake and S.~P. Trivedi, \emph{{Jackiw-Teitelboim gravity in the second order formalism}}, \href{https://doi.org/10.1007/JHEP10(2021)204}{\emph{JHEP} {\bfseries 10} (2021) 204} [\href{https://arxiv.org/abs/2101.00596}{{\ttfamily 2101.00596}}].

\bibitem{Blacker:2025zca}
M.~J. Blacker, A.~Castro, W.~Sybesma and C.~Toldo, \emph{{Quantum corrections to the path integral of near extremal de Sitter black holes}},  \href{https://arxiv.org/abs/2503.14623}{{\ttfamily 2503.14623}}.

\bibitem{Shi:2025amq}
X.~Shi and G.~J. Turiaci, \emph{{The phase of the gravitational path integral}}, \href{https://doi.org/10.1007/JHEP07(2025)047}{\emph{JHEP} {\bfseries 07} (2025) 047} [\href{https://arxiv.org/abs/2504.00900}{{\ttfamily 2504.00900}}].

\bibitem{Bobev:2022lcc}
N.~Bobev, T.~Hertog, J.~Hong, J.~Karlsson and V.~Reys, \emph{{Microscopics of de Sitter Entropy from Precision Holography}}, \href{https://doi.org/10.1103/PhysRevX.13.041056}{\emph{Phys. Rev. X} {\bfseries 13} (2023) 041056} [\href{https://arxiv.org/abs/2211.05907}{{\ttfamily 2211.05907}}].

\bibitem{Astefanesei:2003gw}
D.~Astefanesei, R.~B. Mann and E.~Radu, \emph{{Reissner-Nordstrom-de Sitter black hole, planar coordinates and dS / CFT}}, \href{https://doi.org/10.1088/1126-6708/2004/01/029}{\emph{JHEP} {\bfseries 01} (2004) 029} [\href{https://arxiv.org/abs/hep-th/0310273}{{\ttfamily hep-th/0310273}}].

\bibitem{Montero:2019ekk}
M.~Montero, T.~Van~Riet and G.~Venken, \emph{{Festina Lente: EFT Constraints from Charged Black Hole Evaporation in de Sitter}}, \href{https://doi.org/10.1007/JHEP01(2020)039}{\emph{JHEP} {\bfseries 01} (2020) 039} [\href{https://arxiv.org/abs/1910.01648}{{\ttfamily 1910.01648}}].

\bibitem{Aalsma:2023mkz}
L.~Aalsma, J.~P. van~der Schaar and M.~R. Visser, \emph{{Extremal Black Hole Decay in de Sitter Space}},  \href{https://arxiv.org/abs/2311.13742}{{\ttfamily 2311.13742}}.

\bibitem{Banihashemi:2022jys}
B.~Banihashemi and T.~Jacobson, \emph{{Thermodynamic ensembles with cosmological horizons}}, \href{https://doi.org/10.1007/JHEP07(2022)042}{\emph{JHEP} {\bfseries 07} (2022) 042} [\href{https://arxiv.org/abs/2204.05324}{{\ttfamily 2204.05324}}].

\bibitem{Spradlin:2001pw}
M.~Spradlin, A.~Strominger and A.~Volovich, \emph{{Les Houches lectures on de Sitter space}},  in \emph{{Les Houches Summer School: Session 76: Euro Summer School on Unity of Fundamental Physics: Gravity, Gauge Theory and Strings}}, pp.~423--453, 10, 2001, \href{https://arxiv.org/abs/hep-th/0110007}{{\ttfamily hep-th/0110007}}.

\bibitem{Klemm:2004mb}
D.~Klemm and L.~Vanzo, \emph{{Aspects of quantum gravity in de Sitter spaces}}, \href{https://doi.org/10.1088/1475-7516/2004/11/006}{\emph{JCAP} {\bfseries 11} (2004) 006} [\href{https://arxiv.org/abs/hep-th/0407255}{{\ttfamily hep-th/0407255}}].

\bibitem{Anninos:2012qw}
D.~Anninos, \emph{{De Sitter Musings}}, \href{https://doi.org/10.1142/S0217751X1230013X}{\emph{Int. J. Mod. Phys. A} {\bfseries 27} (2012) 1230013} [\href{https://arxiv.org/abs/1205.3855}{{\ttfamily 1205.3855}}].

\bibitem{Banihashemi:2022htw}
B.~Banihashemi, T.~Jacobson, A.~Svesko and M.~Visser, \emph{{The minus sign in the first law of de Sitter horizons}}, \href{https://doi.org/10.1007/JHEP01(2023)054}{\emph{JHEP} {\bfseries 01} (2023) 054} [\href{https://arxiv.org/abs/2208.11706}{{\ttfamily 2208.11706}}].

\bibitem{Cho:2007we}
J.-H. Cho and S.~Nam, \emph{{The Entropy function for the black holes of Nariai class}}, \href{https://doi.org/10.1088/1126-6708/2008/03/027}{\emph{JHEP} {\bfseries 03} (2008) 027} [\href{https://arxiv.org/abs/0711.2514}{{\ttfamily 0711.2514}}].

\bibitem{GonzalezLezcano:2023cuh}
A.~Gonz\'alez~Lezcano, I.~Jeon and A.~Ray, \emph{{Supersymmetric localization: \ensuremath{\mathscr{N}} = (2) theories on S$^{2}$ and AdS$_{2}$}}, \href{https://doi.org/10.1007/JHEP07(2023)056}{\emph{JHEP} {\bfseries 07} (2023) 056} [\href{https://arxiv.org/abs/2302.10370}{{\ttfamily 2302.10370}}].

\bibitem{GonzalezLezcano:2023uar}
A.~Gonz\'alez~Lezcano, I.~Jeon and A.~Ray, \emph{{Supersymmetry and complexified spectrum on Euclidean AdS2}}, \href{https://doi.org/10.1103/PhysRevD.108.045018}{\emph{Phys. Rev. D} {\bfseries 108} (2023) 045018} [\href{https://arxiv.org/abs/2305.12925}{{\ttfamily 2305.12925}}].

\bibitem{GonzalezLezcano:2024rsi}
A.~Gonz\'alez~Lezcano, I.~Jeon and A.~Ray, \emph{{Supersymmetric spectrum for vector multiplet on Euclidean AdS$_{2}$}}, \href{https://doi.org/10.1007/JHEP08(2024)139}{\emph{JHEP} {\bfseries 08} (2024) 139} [\href{https://arxiv.org/abs/2404.18376}{{\ttfamily 2404.18376}}].

\bibitem{Correa:2019rdk}
D.~H. Correa, V.~I. Giraldo-Rivera and G.~A. Silva, \emph{{Supersymmetric mixed boundary conditions in AdS$_{2}$ and DCFT$_{1}$ marginal deformations}}, \href{https://doi.org/10.1007/JHEP03(2020)010}{\emph{JHEP} {\bfseries 03} (2020) 010} [\href{https://arxiv.org/abs/1910.04225}{{\ttfamily 1910.04225}}].

\bibitem{Hawking:1976ja}
S.~W. Hawking, \emph{{Zeta Function Regularization of Path Integrals in Curved Space-Time}}, \href{https://doi.org/10.1007/BF01626516}{\emph{Commun. Math. Phys.} {\bfseries 55} (1977) 133}.

\bibitem{GotoThesis}
K.~Goto, \emph{{Near horizon physics of charged black holes and the Jackiw-Teitelboim gravity}}, Ph.D. thesis, Tokyo U.

\end{thebibliography}\endgroup

\end{document}